\documentclass[a4paper,11pt]{article}
\pdfoutput=1
\usepackage{jcappub}
\usepackage{subfigure}
\usepackage{graphicx}
\usepackage{color}
\usepackage{makecell}
\usepackage{afterpage}
\usepackage[dvipsnames]{xcolor}
\usepackage{multirow}
\usepackage{cleveref}
\usepackage[normalem]{ulem}
\usepackage{xspace}
\usepackage{booktabs}
\usepackage{centernot}
\usepackage{mathtools}
\usepackage{stmaryrd}
\usepackage[utf8]{inputenc}
\usepackage{csquotes}
\usepackage{placeins}
\usepackage{titlesec}
\usepackage{xcolor}

\title{Pseudoscalar sterile neutrino self-interactions in light of Planck, SPT and ACT data}

\author[a,b]{Mattia Atzori Corona,}
\author[c,h]{Riccardo Murgia,}
\author[b]{Matteo Cadeddu,}
\author[d,e]{Maria Archidiacono,}
\author[f]{Stefano Gariazzo,}
\author[f]{Carlo Giunti,}
\author[g]{Steen Hannestad}

\affiliation[a]{Dipartimento di Fisica, Universit\`a degli Studi di Cagliari,
Complesso Universitario di Monserrato - S.P. per Sestu Km 0.700, 09042 Monserrato, Italy}
\affiliation[b]{INFN -- Istituto Nazionale di Fisica Nucleare, Sezione di Cagliari,
Complesso Universitario di Monserrato - S.P. per Sestu Km 0.700, 09042 Monserrato, Italy}
\affiliation[c]{LUPM -- Laboratoire Univers \& Particules de Montpellier, CNRS \& Universit\'e de Montpellier (UMR-5299), Place Eugène Bataillon, F-34095 Montpellier Cedex 05, France}
\affiliation[d]{Universit\`a degli Studi di Milano, via G. Celoria 16, 20133 Milano, Italy}
\affiliation[e]{INFN -- Istituto Nazionale di Fisica Nucleare, Sezione di Milano,
Via G. Celoria 16, 20133 Milano, Italy}
\affiliation[f]{INFN -- Istituto Nazionale di Fisica Nucleare, Sezione di Torino, Via P. Giuria 1, 10125 Turin, Italy}
\affiliation[g]{Department of Physics and Astronomy, Aarhus University, DK-8000 Aarhus C, Denmark}
\affiliation[h]{Gran Sasso Science Institute (GSSI), I-67100 L’Aquila, Italy}

\emailAdd{mattia.atzori.corona@ca.infn.it}
\emailAdd{riccardo.murgia@gssi.it}
\emailAdd{matteo.cadeddu@ca.infn.it}
\emailAdd{maria.archidiacono@unimi.it}
\emailAdd{gariazzo@to.infn.it}
\emailAdd{carlo.giunti@to.infn.it}
\emailAdd{steen@phys.au.dk}

\abstract{
We reassess the viability of a cosmological model including a fourth additional sterile neutrino species that self-interacts through a new pseudoscalar degree of freedom.
We perform a series of extensive analyses fitting various combinations of cosmic microwave background (CMB) data from \emph{Planck}, the Atacama Cosmology Telescope (ACT) and the South Pole Telescope (SPT), both alone and {in combination} with Baryon Acoustic Oscillation (BAO) and Supernova Ia (SnIa) observations.
We show that the scenario under study, although capable to resolve the Hubble tension without worsening the so-called $S_8$ tension about the growth of cosmic structures, is severely constrained by high-multipole polarization data from both \emph{Planck} and SPT.
Intriguingly, when trading \emph{Planck} TE-EE data for those from ACT, we find a $\gtrsim 3 \sigma$ preference for a non-zero sterile neutrino mass,~$m_s=3.6^{+1.1}_{-0.6}$~eV (68\%~C.L.), compatible with the range suggested by longstanding short-baseline (SBL) anomalies in neutrino oscillation experiments.
The pseudoscalar model provides indeed a better fit to ACT data compared to $\Lambda$CDM ($\Delta\chi^2 \simeq -5$, $\Delta \rm{AIC}=-1.3$), although in a combined analysis with $Planck$ the $\Lambda$CDM model is still favoured, as the preference for a non-zero sterile neutrino mass is mostly driven by ACT favouring a higher value for the primordial spectral index $n_s$ with respect to $Planck$.
We show {that} the mild tension between \emph{Planck} and ACT {is} due to the different pattern in the TE and EE power spectra on multipoles between $350 \lesssim \ell \lesssim 1000$.
We also check the impact of marginalizing over the gravitational lensing information in \emph{Planck} data, showing that the model does not solve the CMB lensing anomaly.
Future work including higher precision data from current and upcoming CMB ground-based experiments will be crucial to test these results.
}

\begin{document}

\maketitle

\section{Introduction}\label{sec:intro}

The $\Lambda$ Cold Dark Matter ($\Lambda$CDM) {cosmological} model has been proven to provide an excellent fit both to early-universe observations, such as the cosmic microwave background (CMB)~\cite{Planck:2018vyg}, and to late-universe measurements, such as large-scale structure data (LSS) data~\cite{Alam:2016hwk,eBOSS:2020yzd}.
Nonetheless, within such a standard framework there exist a few tensions, most notably between the early-universe, indirect
determinations of the Hubble parameter $H_0$ and of the parameter $S_8\equiv\sigma_8(\Omega_m/0.3)^{0.5}$ -- where $\sigma_8$ is the root mean square of matter fluctuations on a 8 $h^{-1}$Mpc scale, and $\Omega_m$ is the total matter abundance -- compared to their direct, low-redshift measurements, {respectively} from calibrated SnIa as a cosmic distance ladder~\cite{Riess:2019cxk,Freedman:2019jwv,Yuan:2019npk,Cerny:2020inj,Soltis:2020gpl,Riess:2020fzl,Dainotti:2021pqg,Blakeslee:2021rqi,Anand:2021sum} and weak gravitational lensing~\cite{Hildebrandt:2018yau,HSC:2018mrq,Joudaki:2019pmv,Heymans:2020gsg,DES:2021wwk,Gatti:2021uwl}.
In spite of meticulous attempts to check for possibly unaccounted systematics at play in the local estimates of such parameters~\cite{Rigault:2014kaa,NearbySupernovaFactory:2018qkd,DES:2021wwk}, both the so-called Hubble and growth tensions persist, and have nowadays reached about the 5 and 3 $\sigma$ level, respectively~\cite{Knox:2019rjx,Verde:2019ivm,DiValentino:2020zio,Perivolaropoulos:2021jda,Freedman:2021ahq}.

While a resolution to the growth tension can be achieved in a number of models departing from $\Lambda$CDM at late times without affecting pre-recombination physics (e.g.~\cite{Camera:2017tws,DiValentino:2017zyq,Lambiase:2018ows,Abellan:2020pmw,DiValentino:2020vvd}), purely late-time {explanations of} the Hubble tension have been shown to be the least viable, due to the SnIa and BAO constraints at $z \lesssim 2$~\cite{Bernal:2016gxb,Jedamzik:2020zmd}.
However, {currently there is an open debate on the possibility that} introducing new physics in the pre-recombination era could resolve the Hubble tension without spoiling other bounds or exacerbating the growth tension~\cite{Knox:2019rjx,Jedamzik:2020zmd,Haridasu:2020pms,DiValentino:2021izs,Vagnozzi:2021gjh}.
It seems indeed that, in order to fully restore cosmological concordance, it might be necessary to modify both the early-universe physics, e.g.\ by reducing the sound horizon at recombination to accommodate a higher $H_0$, and the late-universe physics, to decrease the amplitude of matter fluctuations on scales $k\sim 0.1-1 ~h$/Mpc~\cite{DiValentino:2021izs,Schoneberg:2021qvd}.
One possibility, motivated by particle physics~\cite{Abazajian:2012ys,Giunti:2019aiy} is the introduction of a light sterile neutrino species, namely a singlet state under the SU(2)$_L\otimes$U(1)$_Y$ electroweak gauge group.
This additional degree of freedom {would} not interact via any of the fundamental interactions of the Standard Model, {but would oscillate with the active neutrino species}. 

The existence of a sterile neutrino {with a mass in the eV range} is motivated by the fact that it might provide an explanation to long-standing anomalies in short-baseline (SBL) neutrino oscillation experiments.
{These include the anomalous appearance of events measured by} the LSND~\cite{LSND:2001aii} and MiniBooNE~\cite{MiniBooNE:2018esg, MiniBooNE:2020pnu} experiments,
{and the anomalous disappearance of electron (anti)neutrinos detected} by several observations measuring the electron antineutrino flux from nuclear reactors \cite{Mention:2011rk} and in the calibration of the GALLEX~\cite{Kaether:2010ag} and SAGE~\cite{SAGE:2009eeu} gallium solar neutrino experiments \cite{Giunti:2006bj,Giunti:2010zu}
(see Refs.\ \cite{Gariazzo:2015rra,Boser:2019rta,Giunti:2019aiy} for a full list of references).
Although the sterile neutrino hypothesis was claimed to provide an explanation to all these anomalies at once \cite{Kopp:2013vaa,Gariazzo:2015rra},
the tension between appearance and disappearance channels
has increased to a very strong level in the recent years~\cite{Dentler:2018sju,Gariazzo:2017fdh,Diaz:2019fwt}.
Moreover, recent re-analyses of the reactor data~\cite{Berryman:2020agd,Giunti:2021kab}
have reduced the significance of the reactor antineutrino anomaly.
On the other hand,
the Gallium anomaly,
which was reduced by the shell model reevaluation of the cross section in Ref.~\cite{Kostensalo:2019vmv},
has been recently revived by the result of the BEST experiment~\cite{Barinov:2021asz}
(see also the discussions in Refs.~\cite{Barinov:2021mjj,Giunti:2021kab,Berryman:2021yan}).
\looseness=-1
Considering the $\nu_\mu\to\nu_e$ appearance channel,
the new results of the MicroBooNE experiment \cite{MicroBooNE:2021jwr,MicroBooNE:2021nxr,MicroBooNE:2021rmx}
disfavour the sterile neutrino interpretation of the MiniBooNE anomaly as an electron neutrino appearance from a muon neutrino beam
(see, however, the discussion in Ref.~\cite{Arguelles:2021meu}).
It is interesting that a recent analysis shows a 2.2$\sigma$ preference for a sterile neutrino mass in the eV scale if the MicroBooNE data are interpreted in terms of electron neutrino disappearance~\cite{Denton:2021czb}. 

CMB and LSS observations strongly constrain the simplest scenario where the new sterile neutrino component is a non-interacting and free-streaming species \cite{Gariazzo:2019gyi,Hagstotz:2020ukm,Gariazzo:2018zho,Gariazzo:2016ehl,Gariazzo:2015rra}.
In such a minimal scenario, it is therefore very unlikely to find a common resolution to SBL anomalies and cosmological tensions.
That is why several models beyond the simple non-interacting case have been proposed in the literature, in particular scenarios where the sterile neutrinos are coupled through new interactions~\cite{Dasgupta:2013zpn,Hannestad:2013ana,Archidiacono:2014nda,Kreisch:2019yzn}.

In this work, we focus on a specific self-interacting sterile neutrino scenario -- introduced in Refs.~\cite{Archidiacono:2014nda,Archidiacono:2015oma} and subsequently tested against cosmological and SBL data in Refs.~\cite{Archidiacono:2016kkh,Archidiacono:2020yey} -- where a light massive sterile neutrino species self-interacts through the exchange of a new massless pseudoscalar degree of freedom.
This model induces a radically different phenomenology compared to the non-interacting case, because the sterile neutrino is not a free-streaming species.
In fact, due to its self-interaction, it can be treated as a single tightly coupled fluid
{together with the pseudoscalar}.
Moreover, the rapid pair-annihilation and disappearance when the temperature drops below its mass prevents the pseudoscalar model from violating constraints from LSS observations~\cite{Archidiacono:2014nda,Archidiacono:2015oma}.
Although this scenario can readily ease the Hubble tension, a non-zero sterile neutrino mass, mildly favoured by \emph{Planck} CMB temperature data~\cite{Archidiacono:2016kkh}, appears to be very tightly constrained when high-multipole \emph{Planck} polarization data are added to the analysis~\cite{Archidiacono:2020yey}.

Let us now introduce another anomaly characterizing the standard cosmological framework: the so-called CMB lensing (or ``A$_{\rm lens}$'') anomaly, i.e.\ a residual oscillatory feature in \emph{Planck} data at high multipoles ($1000 \lesssim \ell \lesssim 2000$) compared to the best-fit $\Lambda$CDM prediction~\cite{Calabrese:2008rt,Aghanim:2016sns,Planck:2018vyg,Efstathiou:2019mdh}.
Such a feature can be described as an extra source of smoothing of the acoustic peaks, and modelled via two extra {phenomenological} parameters: $A_{\rm L}^{\rm TTTEEE}$, that controls the amount of smoothing, and $A_{\rm L}^{\rm \phi\phi}$, that re-scales the global amplitude of the lensing potential power spectrum. Both these parameters are predicted to be equal to one within the $\Lambda$CDM model. While the lensing anomaly can be observed in the TT-TE-EE spectra, the amount of gravitational lensing can also be determined directly from the lensing potential power spectrum reconstructed from the CMB four-point correlation function, and {in this case it} is compatible with the $\Lambda$CDM expectation ($A_{\rm L}^{\rm \phi\phi} = 1$).
On the other hand, the case where $A_{\rm L}^{\rm TTTEEE} = 1$ is about $3 \sigma$ away from the $\Lambda$CDM best-fit.
It thus seems that the extra smoothing of the TT-TE-EE peaks cannot be attributed to actual gravitational lensing~\cite{Aghanim:2016sns, Motloch:2018pjy, Motloch:2019gux}.
Furthermore, once marginalizing over the lensing information in \emph{Planck} data, the resulting temperature and polarization power spectra favor a cosmology with a lower $A_s$ and $\omega_{\rm cdm} \equiv \Omega_{\rm cdm}h^2$.
Indeed, these parameters are strongly correlated with the amplitude of the lensing potential power spectrum.
As a consequence, the ``$\Lambda$CDM+A$_{\rm lens}$'' cosmology shows no growth tension and a slightly alleviated Hubble tension.
Moreover, such a cosmology is in better agreement with the $\Lambda$CDM best-fit cosmology reconstructed from data collected by ongoing ground-based CMB experiments at the South Pole Telescope (SPT)~\cite{Henning:2017nuy,Chudaykin:2020acu} and Atacama Cosmology Telescope (ACT)~\cite{Aiola:2020azj,Choi:2020ccd} (see, e.g.~Ref.~\cite{Handley:2020hdp}).
No departure from the case where $A_{\rm L}^{\rm TTTEEE} = 1$ is indeed {preferred} by SPT and ACT.
The introduction of $A_{\rm{L}}^{\rm{TTTEEE}}$ and $A_{\rm{L}}^{\phi\phi}$ modify the correlation between cosmological parameters both in the presence of an additional free-streaming component, as in the non-interacting sterile neutrino model, and in the pseudoscalar scenario, where the sterile neutrino component behaves like a coupled fluid rather than a free-streaming species. It is thus worth studying whether such a multi-parameter degeneracy can alleviate the lensing anomaly in each of these two scenarios, as well as investigating the impact on the sterile neutrino sector parameters.

In light of all these considerations, our goal is to test the robustness of the limits obtained in Ref.~\cite{Archidiacono:2020yey} under the following changes in the CMB data analysis:
\begin{itemize}
\item trading the high-multipole TE-EE data from {\emph{Planck}} for those from SPT\footnote{We made use of the SPTpol data and likelihood, being the only publicly available likelihood for SPT data that can be interfaced with the MCMC sampler used in this work. A more recent data-set, SPT-3G~\cite{SPT-3G:2021eoc}, was released by the SPT Collaboration when our work was already in an advanced stage. We leave an analysis of the pseudoscalar scenario with SPT-3G data for a future work.}, as in Refs.~\cite{Chudaykin:2020acu,Chudaykin:2020igl,Abellan:2021bpx};
\item trading the high-multipole TE-EE data from {\emph{Planck}} for those from ACT, as in Refs.~\cite{Lin:2020jcb,Abellan:2021bpx,Galli:2021mxk,Hill:2021yec,Poulin:2021bjr}.
\item introducing two additional free parameters, $A^{\rm TTTEEE}_{\rm L}$ and $A^{\phi\phi}_{\rm L}$, the former capturing the impact of gravitational lensing on the TT-TE-EE spectra, the latter globally re-scaling the amplitude of the lensing potential power spectrum, in order to marginalize over the lensing anomaly in \emph{Planck} data, as in Refs.~\cite{Simard:2017xtw,Wu:2019hek,Murgia:2020ryi,Abellan:2021bpx};
\end{itemize}

This work is structured as follows: in Sec.~\ref{sec:model} we briefly outline the theoretical framework under study; in Sec.~\ref{sec:data} we discuss the data-sets that we have considered and the methodology that we have adopted; in Sec.~\ref{sec:results} we present our results; Sec.~\ref{sec:act2} is dedicated to a deeper scrutiny of the analyses with ACT data; in Sec.~\ref{sec:sbl} we briefly discuss the compatibility of our cosmological results with up-to-date constraints from SBL neutrino oscillation experiments; finally, in Sec.~\ref{sec:concl} we draw our conclusions and outline future perspectives.

\section{The pseudoscalar sterile neutrino self-interaction model}\label{sec:model}

The theoretical framework under investigation -- introduced in Ref.~\cite{Archidiacono:2014nda} and subsequently reassessed in light of different experimental constraints in Refs.~\cite{Archidiacono:2014nda,Archidiacono:2015oma,Archidiacono:2016kkh,Archidiacono:2020yey} -- is a cosmological scenario where a sterile neutrino species couples to an effectively massless pseudoscalar degree of freedom.
In this Section we briefly recall its basic features and phenomenological parametrisation.

The Lagrangian term describing the coupling between sterile neutrinos and the
new pseudoscalar field $\phi$, with mass $m_\phi \ll 1\;\rm{eV}$, is given by:
\begin{equation}
    \mathcal{L}\sim g_s \phi \bar{\nu}_4\gamma^5 \nu_4,
\end{equation}
where $\nu_4$ is the fourth -- mainly sterile -- neutrino mass state, and $g_s$ is the coupling constant that characterizes the intensity of the interaction.
The new interaction is also partly felt by active neutrinos, although {in this case its strength is} suppressed by the {active-sterile} mixing angle. 
If the dimensionless coupling is larger than $g_s\sim 10^{-6}$, the production of sterile neutrinos, {which causes} an increase of $N_{\rm{eff}}$, is delayed until the time of active neutrino decoupling {when active-sterile oscillations are not effective anymore. This moment also} roughly coincides with the onset of BBN, allowing to evade the bounds from the latter~\cite{Schoneberg:2019wmt}.
After neutrinos decouple from the plasma, the energy in the neutrino-pseudoscalar sector is redistributed by oscillations so that the sterile plus pseudoscalar sector ends up with a fraction of 11/32 of the total energy density, while the remaining fraction 21/32 goes to the active sector.
After that, provided that $g_s\gtrsim 10^{-6}$, the active {neutrinos} and the sterile-pseudoscalar components are completely decoupled and do not exchange neither energy nor momentum.
The sterile neutrinos become very strongly coupled with the pseudoscalar field and the system can be treated as a single fluid with a well-defined energy density and equation of state.
As soon as sterile neutrinos become non-relativistic, they annihilate into $\phi$, {which is effectively massless}, so that this mechanism allows evading limits on the neutrino mass arising from LSS.
For these reasons, whereas the non-interacting sterile neutrino parameter space is strongly constrained by the aforementioned cosmological probes, the pseudoscalar model could potentially allow to reconcile O(eV) sterile neutrinos with cosmology.
Given that the value of $g_s$ {has an unique correspondence with} the effective number of relativistic degrees of freedom $N_{\rm eff}$ (see Fig.1 from Ref.~\cite{Archidiacono:2014nda}), the pseudoscalar model features only two additional free parameters: the sterile neutrino mass $m_s$, and its contribution to the effective number of relativistic degrees of freedom $\Delta N_{\rm eff}$.
We address the reader to the aforementioned Refs.~\cite{Archidiacono:2014nda,Archidiacono:2015oma,Archidiacono:2016kkh,Archidiacono:2020yey} for a comprehensive description of the model.

\section{Methods and data}\label{sec:data}

We test the pseudoscalar self-interacting sterile neutrino model on a {number} of cosmological observations, by means of a set of comprehensive {Markov Chain Monte Carlo} (MCMC) analyses with the \texttt{MontePython-v3}\footnote{\url{https://github.com/brinckmann/montepython_public}} sampler \cite{Audren:2012wb,Brinckmann:2018cvx},
interfaced with a modified version of the numerical Einstein-Boltzmann solver \texttt{CLASS}~\cite{Blas:2011rf,Archidiacono:2020yey}.
{We consider} various combinations of the following data-sets:
\begin{itemize}
\item the low-$\ell$ CMB TT, EE ($ \ell < 30$), the high-$\ell$ TT, TE, EE ($ 30 \leq \ell \leq 2500$) data~\cite{Planck:2018vyg}, and the gravitational lensing potential reconstruction ($ 8 \leq \ell \leq 400$)~\cite{Planck:2018lbu} from {\emph{Planck}} 2018; 
\item the high-$\ell$ CMB EE and TE ($50 \leq \ell \leq 8000$)~\cite{Henning:2017nuy} measurements, and the reconstructed gravitational lensing potential ($100 \leq \ell \leq 8000$)~\cite{Bianchini:2019vxp} from the 500deg SPTpol 
survey~\cite{Chudaykin:2020acu};
\item the high-$\ell$ CMB TT, EE and TE ($ 350 \leq \ell \leq 4125$) data from the DR4 of the ACT survey~\cite{Aiola:2020azj,Choi:2020ccd};
\item the BAO measurements from 6dFGS at $z=0.106$~\cite{Beutler:2011hx}, SDSS DR7 at $z=0.15$~\cite{Ross:2014qpa}, BOSS DR12 at $z=0.38, 0.51$ and $0.61$~\cite{Alam:2016hwk}, and the joint constraints from eBOSS DR14 Ly-$\alpha$ auto-correlation at $z=2.34$~\cite{Agathe:2019vsu} and cross-correlation at $z=2.35$~\cite{Blomqvist:2019rah};
\item the measurements of the growth function $f\sigma_8(z)$ (FS) from the CMASS and LOWZ galaxy samples of BOSS DR12 at $z = 0.38$, $0.51$, and $0.61$~\cite{Alam:2016hwk};
\item the Pantheon SnIa catalogue, spanning redshifts $0.01 < z < 2.3$~\cite{Scolnic:2018rjj}.
\end{itemize}

Our baseline cosmology is described by the standard set of six $\Lambda$CDM parameters, namely the baryon and cold dark matter physical energy densities ($\omega_b$, $\omega_{\rm{cdm}}$), the angular size of the sound horizon at recombination ($\theta_s$), the tilt and amplitude of the primordial power spectrum ($n_s$, $A_s$), and the optical depth at reionization ($\tau_{\rm{reio}}$). The pseudoscalar scenario is fully characterized by two additional parameters describing the sterile neutrino sector, i.e.~the sterile neutrino mass, $m_s$, and its contribution to the effective number of relativistic degrees of freedom, $\Delta N_{\rm eff}$.
We adopt flat priors on all the parameters\footnote{When making use of ACT data, we adopt a Gaussian prior on the optical depth to reionization, $\tau_{\rm reio} = 0.06 \pm 0.01$, as suggested by the ACT collaboration, to overcome the lack of information on low multipoles. We verified that including low-$\ell$ EE data from \textit{Planck} rather than adopting such a prior choice on $\tau_{\rm reio}$ does not affect our conclusions.}, and we assume massless active neutrinos {for simplicity}.
We consider MCMC chains to be converged when the Gelman-Rubin criterion~\cite{Gelman:1992zz} satisfies $R-1<0.03$.
We analyze the results with the \texttt{Getdist} python package\footnote{\url{https://getdist.readthedocs.io/}}~\cite{Lewis:2019xzd}, and extract best-fit parameters making use of the {\texttt{}{Minuit}} algorithm~\cite{JAMES1975343} through the {\texttt{iMinuit}} python package\footnote{\url{https://iminuit.readthedocs.io/}}.
We primarily focus on the results in each of the CMB-only set-ups that we have considered, and then discuss the impact of adding BAO and SnIa data to the analyses.

\section{Results}\label{sec:results}

\begin{figure}[ht]
 \centering
   \includegraphics[scale=0.4]{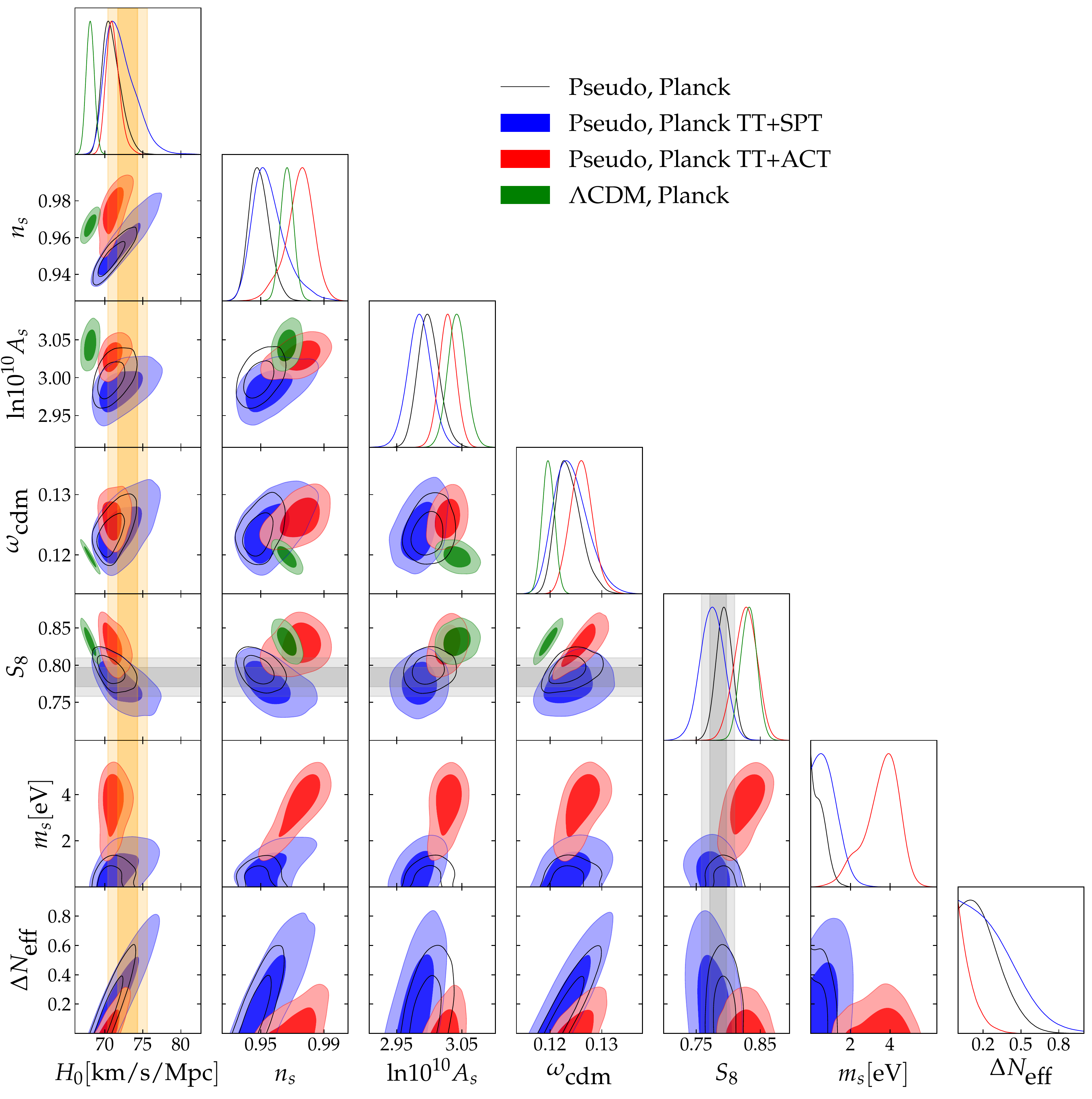} %
  \caption{Posterior distributions of the cosmological parameters in the $\Lambda$CDM and pseudoscalar (Pseudo) model for different CMB data-sets. The orange and gray bands represent the direct measurements (1$\sigma$ and 2$\sigma$ confidence regions) of $H_0$ and $S_8$, from cosmic distance ladder~\cite{Riess:2020fzl} and weak lensing observations~\cite{Gatti:2021uwl}, respectively.
  }
  \label{Fig:pseudo_LCDM_Planck_SPT_ACT}
\end{figure}

Let us first briefly discuss the current constraints on the pseudoscalar model from {\it Planck} data, that are shown as empty black contours in Fig.~\ref{Fig:pseudo_LCDM_Planck_SPT_ACT}, together with the limits from the other data combinations that will be discussed in the following Subsections. As a reference, we also report \textit{Planck} limits on the $\Lambda$CDM model and the bands representing the direct measurement of the Hubble constant from cosmic distance ladder ($H_0=73.0^{+1.3}_{-1.3}$ km/s/Mpc~\cite{Riess:2020fzl}) and $S_8$ from weak lensing observations ($S_8=0.784^{+0.013}_{-0.013}$~\cite{Gatti:2021uwl}).
The corresponding mean values and $\pm 1 \sigma$~C.L.\ are reported in the first column of Tab.~\ref{tab:bf1}.
We substantially confirm the capability of the model under study to provide higher values for $H_0$ with respect to $\Lambda$CDM, mostly {because of} the well-known degeneracy between the latter parameter and the extra-contribution to the effective number of relativistic degrees of freedom $\Delta N_{\rm eff}$.
It is also interesting to notice that the pseudoscalar scenario predicts lower values for the amplitude and tilt of the primordial power spectrum, $A_s$ and $n_s$, resulting in a lower value for $S_8$, so that the model could in principle resolve both the Hubble tension and the growth tension.
However, as already extensively discussed in Ref.~\cite{Archidiacono:2020yey}, \textit{Planck} polarization data on multipoles larger than $\ell = 30$ severely constrain the possibility of a non-zero sterile neutrino mass.
The degradation of the global fit with respect to $\Lambda$CDM ($\Delta \chi^2 \simeq 13$) is indeed fully driven by a poor fit to that subset of \textit{Planck} data (see App.~\ref{app:bfchi2}, where we report all individual $\chi^2$'s).


\begin{table}[t]
\centering
\scalebox{0.73}{
  \begin{tabular}{|l||c|c|c|c|c|c|c|}
    \hline
& \multicolumn{4}{c|}{{Pseudo vs CMB}}\\ \hline\hline
Parameter & \emph{Planck} & \emph{Planck} w. $A_\textrm{lens}$ & \emph{Planck} TT+SPT & \emph{Planck} TT+ACT\\
\hline \hline

100 $\omega_b$ & $2.269_{-0.025}^{+0.025}$  & $2.319_{-0.032}^{+0.030}$ & $2.272_{-0.036}^{+0.032}$ & $2.216_{-0.022}^{+0.018}$
\\
$\omega_{\rm{cdm}}$ & $0.1237_{-0.0029}^{+0.0020}$  & $0.1223_{-0.0035}^{+0.0027}$ & $0.1239_{-0.0038}^{+0.0026}$ & $0.1261_{-0.0023}^{+0.0022}$
\\
100 $\theta_s$ & $1.04530_{-0.00043}^{+0.00040}$  & $1.04530_{-0.00046}^{+0.00043}$ & $1.04510_{-0.00054}^{+0.00049}$ & $1.04670_{-0.00042}^{+0.00049}$ 
\\
$\textrm{ln}10^{10}A_s$ &  $2.999_{-0.018}^{+0.015}$  & $2.974_{-0.018}^{+0.020}$ & $2.985_{-0.018}^{+0.018}$ & $3.028_{-0.013}^{+0.013}$ 
\\
$n_s$ & $0.9491_{-0.0073}^{+0.0057}$  & $0.9634_{-0.0096}^{+0.0079}$ & $0.955_{-0.013}^{+0.007}$  & $0.975_{-0.007}^{+0.010}$ 
\\
$\tau_{\textrm{reio}}$ & $0.0589_{-0.0087}^{+0.0073}$ & $0.0506_{-0.0079}^{+0.0089}$ & $0.0522_{-0.0078}^{+0.0082}$  & $0.0550_{-0.0062}^{+0.0060}$
\\
$m_s$ [eV] & $<1.1$  & $<1.0$ & $<1.9$ & $3.6_{-0.6}^{+1.1}$
\\
$\Delta N_{\textrm{eff}}$ & $<0.50$   & $0.37_{-0.25}^{+0.17}$ & $<0.67$ & $<0.25$
\\
$A^{\phi\phi}_{\rm{L}}$ & & $1.132_{-0.048}^{+0.044}$ & &
\\
$A^{\rm{TTTEEE}}_{\rm{L}}$ & & $1.241_{-0.080}^{+0.073}$  & &
\\
\hline \hline
$H_0$ [km/s/Mpc] & $71.0_{-1.5}^{+1.0}$  & $73.8_{-2.1}^{+1.6}$ & $72.1_{-2.4}^{+1.4}$ & $71.1_{-1.0}^{+0.8}$
\\
$S_8$ & $0.794_{-0.013}^{+0.013}$ & $0.744_{-0.019}^{+0.019}$ & $0.775_{-0.020}^{+0.018}$ & $0.828_{-0.018}^{+0.019}$
\\
\hline
\end{tabular}}
\caption{The mean $\pm 1~\sigma$ error ($2~\sigma$ in the case of
upper bounds) of the cosmological parameters from CMB experiments in the pseudoscalar model.}
\label{tab:bf1}
\end{table}

\subsection{Planck TT+SPT}

Let us now shortly discuss the constraints on the pseudoscalar model obtained from a set of CMB data constituted by low-$\ell$ temperature and polarization as well as high-$\ell$ temperature data from {\emph{Planck}}, in combination with high-$\ell$ polarization data from SPT.
The resulting contour plots are shown in blue in Fig.~\ref{Fig:pseudo_LCDM_Planck_SPT_ACT}, while the corresponding mean values and $\pm 1 \sigma$~C.L.\ are reported in the third column of Tab.~\ref{tab:bf1}. From Fig.~\ref{Fig:pseudo_LCDM_Planck_SPT_ACT} we clearly see that the \textit{Planck} TT+SPT contours are very similar to the \textit{Planck}-only ones, albeit with larger uncertainties.
As in the $\Lambda$CDM case~\cite{Henning:2017nuy,Chudaykin:2020acu}, even in the pseudoscalar model the inclusion of SPT data favours low $S_8$ values in agreement with weak lensing observations.
However, we find a degradation of the fit with respect to $\Lambda$CDM, which is mainly driven by a worse fit to \textit{Planck} high-$\ell$ TT data ($\Delta\chi^2\simeq8$), and by a mild degradation in the fit to TE and EE data from SPT ($\Delta\chi^2\simeq3$).

From this analysis we conclude that the SPT data-set used in this work does not provide enough constraining power to significantly impact $Planck$ results. For this reason, very similar conclusions could be drawn from our $Planck$ and $Planck$ TT+SPT analyses. Let us stress that, given the large uncertainties of SPT data, the tight limits on the sterile neutrino sector obtained within this data-combination are strongly driven by $Planck$. As one can see by comparing the black and blue contours of Fig.~\ref{Fig:pseudo_LCDM_Planck_SPT_ACT}, the addition of SPT data actually relaxes the bounds, allowing a larger overlap with the predictions from $Planck$ TT+ACT, that we will extensively discuss in the following Sections. Hence, it will be extremely important to confront the pseudoscalar model with the latest, higher-precision data release from SPT-3G~\cite{SPT-3G:2021eoc}. We leave such a study for a future work.

\subsection{Planck TT+ACT}\label{sec:act1}

We now discuss the results obtained performing a joint analysis of \textit{Planck} low-$\ell$ + high-$\ell$ TT data, combined with the ACT DR4 data-set\footnote{We follow the procedure suggested by the ACT collaboration and truncate multipoles $\ell < 1800$ in the ACT TT data to prevent double counting of modes.}, that are reported in red in Fig.~\ref{Fig:pseudo_LCDM_Planck_SPT_ACT}. The corresponding mean values and $\pm 1 \sigma$~C.L.\ are listed in the last column of Tab.~\ref{tab:bf1}. As it is manifest from both Fig.~\ref{Fig:pseudo_LCDM_Planck_SPT_ACT} and Tab.~\ref{tab:bf1}, the inclusion of ACT data drives $n_s$ towards higher values, both in the $\Lambda$CDM and in the pseudoscalar model. 
However, in the $\Lambda$CDM scenario the predictions from the three different data-sets shown in Fig.~\ref{Fig:pseudo_LCDM_Planck_SPT_ACT} are consistent within 1$\sigma$. On the other hand, driven by the fact that ACT data favour $n_s \sim 1$ -- although with large error bars -- in the pseudoscalar scenario the result from \textit{Planck} TT+ACT ($n_s = 0.975_{-0.007}^{+0.010}$) is roughly $2 \sigma$ larger than what we find in the \textit{Planck} and \textit{Planck} TT+SPT analyses,~i.e.\ $n_s = 0.9491_{-0.0073}^{+0.0057}$ and $0.955_{-0.013}^{+0.007}$, respectively.
Strikingly, in the \textit{Planck} TT+ACT analysis the positive correlation between $n_s$ and $m_s$, as apparent in Fig.~\ref{Fig:pseudo_LCDM_Planck_SPT_ACT}, leads to a preference for a non-zero sterile neutrino mass of $m_s=3.6^{+1.1}_{-0.6}$~eV (68\%~C.L.).
In other words, within $\Lambda$CDM, given the absence of an extra-parameter capable to balance the effect of a higher $n_s$, the difference between \textit{Planck} and ACT predictions translates into a lower $n_s$ -- though still slightly higher than that favoured by \textit{Planck} alone -- and thus a degradation of the fit to ACT data. Conversely, in the pseudoscalar scenario the goodness of the fit to ACT data is not altered by the addition of $Planck$ TT data, there is more room for a relatively high $n_s$, and consequently higher values of $m_s$ are allowed.
We will discuss more in detail the degeneracy between the {tilt of the primordial power spectrum} and the sterile neutrino mass in Sec.~\ref{sec:act2}, where we carry out a more extensive investigation on the constraints reported here, aimed at identifying the range of multipoles where ACT and \emph{Planck} and polarization data are somewhat in tension.

The preference for a larger value of $n_s$ (and $A_s$) also translates into higher values for $S_8$, that {consequently} would be at odds roughly as much as the $\Lambda$CDM prediction {with what is measured by weak lensing surveys}.
The parameter $\Delta N_{\rm eff}$ is severely constrained, as one can also see from the last column of Tab.~\ref{tab:bf1}.
Let us finally note that in the joint \textit{Planck} TT+ACT data-set the global fit is only mildly degraded compared to $\Lambda$CDM ($\Delta \chi^2 \simeq 3$), due to the fact that the pseudoscalar model provides a better fit {of} ACT data {than the $\Lambda$CDM model} ($\Delta \chi^2 \simeq -6$), balancing the worse fit to high-$\ell$ \emph{Planck} TT data ($\Delta \chi^2 \simeq 9$).

 \begin{figure}[h!]
 \centering
   \includegraphics[scale=0.45]{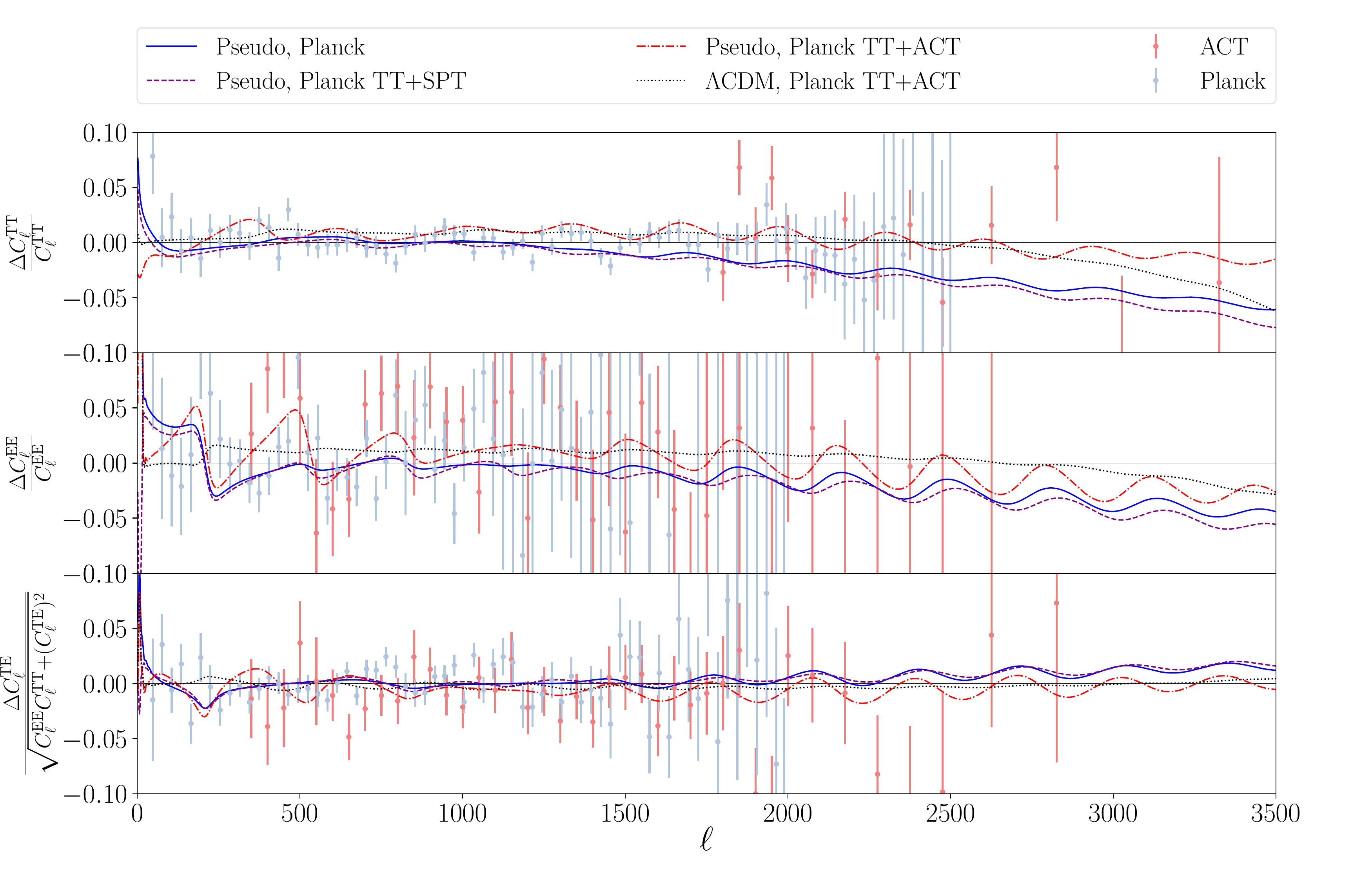}  
  \caption{Residuals in the (lensed) CMB TT, EE and TE power spectrum with respect to the \emph{Planck}-only $\Lambda$CDM best-fit. 
  We also show \emph{Planck} and ACT data-points and error bars.}
  \label{fig: Residual}
\end{figure}

\subsection{Best-fit cosmologies and impact on the CMB power spectra}\label{sec:bf_cosmo}

Let us now explicitly compare the different best-fit CMB angular power spectra from the various analyses that we just discussed.
In Fig.~\ref{fig: Residual} we show the residuals in the CMB TT, EE and TE power spectra for the pseudoscalar model tested against \emph{Planck}, \emph{Planck} TT+SPT, \emph{Planck} TT+ACT data together with \emph{Planck} and ACT data-points and error bars. For comparison, we also show the residuals of the $\Lambda$CDM tested against \emph{Planck} TT+ACT.

In all cases the residuals are computed with respect to the $\Lambda$CDM best-fit from \emph{Planck}-only.
From Fig.~\ref{fig: Residual}, as expected, we first note that the best-fit spectra from the Planck TT+SPT analysis are very similar to those from the \emph{Planck}-only analysis. 
Most importantly, we notice that the \emph{Planck} TT+ACT case, being the only data combination favouring a non-zero value of $m_s$, features indeed the most significant differences in the angular power spectra, especially in TE and EE.
Notice also that the enhanced oscillation pattern around $ \ell \sim 500$ resembles the best-fit power spectrum obtained in the \emph{Planck} TT only analysis by Ref.~\cite{Archidiacono:2020yey}, which indeed also favours a non-zero value for $m_s$ (see Fig.~\ref{Fig:act_new}).
It is of interest to note that the EE best-fit residuals in this multipole range are also very similar to those obtained in very recent analyses against ACT data carried out in the context of models with dark energy at early times~\cite{Lin:2020jcb,Hill:2021yec,Poulin:2021bjr}.
Even in that framework, the mild tension between \emph{Planck} and ACT data on intermediate multipoles translates into hints for new physics beyond $\Lambda$CDM, namely into a slight preference for a non-zero early dark energy component~\cite{Hill:2021yec,Poulin:2021bjr}. Our results, as well as theirs, explicitly demonstrate the importance of examining predictions coming from different CMB data combinations, especially within non-trivial extensions of the $\Lambda$CDM scenario.
The oscillation pattern in the best-fit \emph{Planck} TT+ACT spectra around $ \ell \sim 500$ reflects the trend of ACT data-points. The fact that the latter ones are in mild tension with \emph{Planck} data implies that the $\Lambda$CDM best-fit from \emph{Planck} does not necessarily provide a good fit to ACT data with respect to alternative scenarios, such as the pseudoscalar model under study.
We will discuss in detail sources and implications of the slight inconsistency between \emph{Planck} and ACT in Section~\ref{sec:act2}, also thanks to a series of additional MCMC analyses.

\subsection{Implications for the CMB lensing anomaly}\label{sec:Planck+Alens}

 \begin{figure}[t]
 \centering
   \includegraphics[scale=0.4]{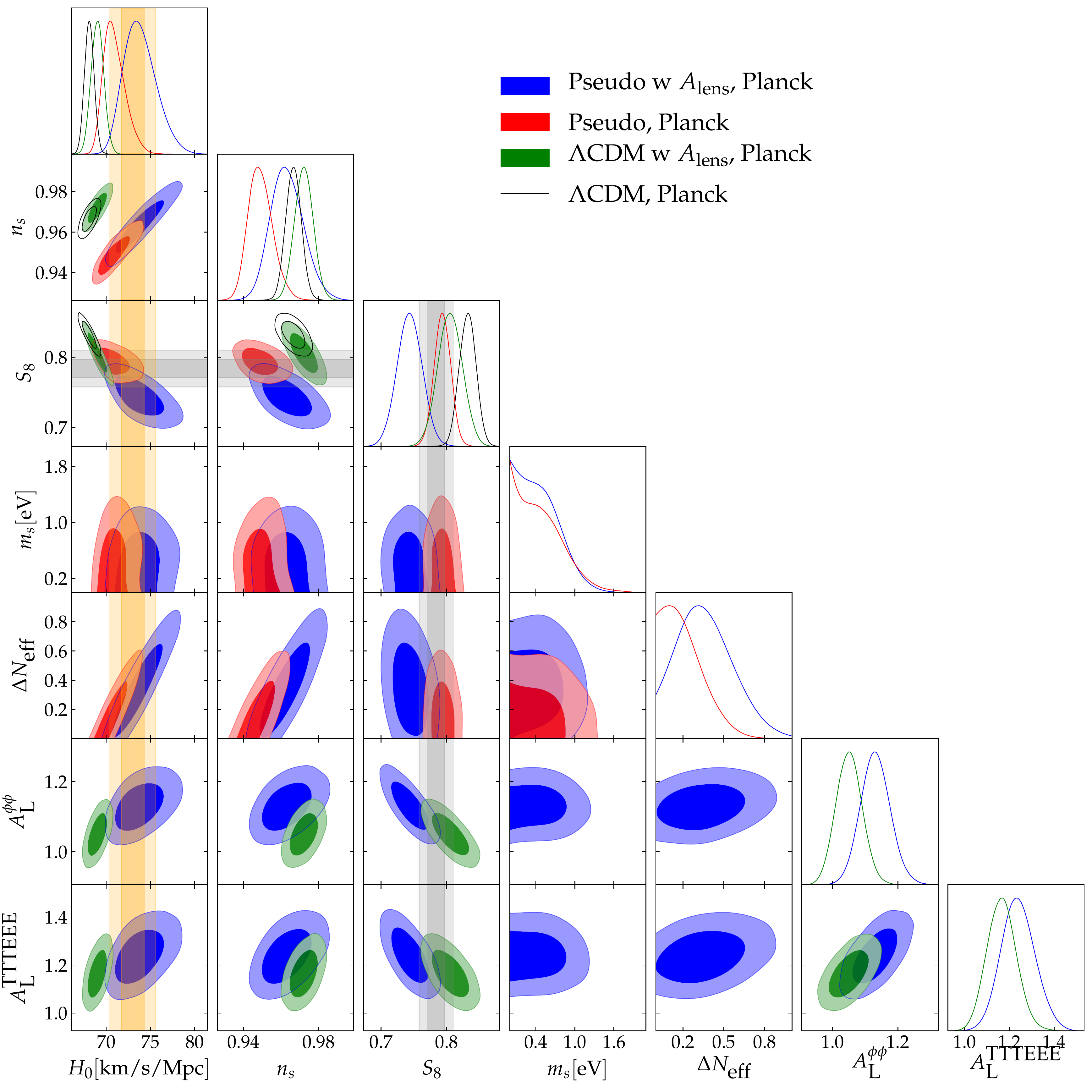}  
  \caption{\looseness=-1 Posterior distributions of the cosmological parameters in the $\Lambda$CDM and pseudoscalar (Pseudo) model tested with \emph{Planck} data, with and without marginalizing over the gravitational lensing information. The orange and gray bands represent the direct measurements (1 $\sigma$ and 2 $\sigma$ confidence regions) of $H_0$ and $S_8$, from cosmic distance ladder~\cite{Riess:2020fzl} and weak lensing data~\cite{Gatti:2021uwl}, respectively.}
  \label{Fig:Pseudo_Alens}
\end{figure}

Let us now focus on the lensing anomaly in \emph{Planck} data, and on studying the robustness of the constraints in the pseudoscalar model when marginalizing over the lensing information in \textit{Planck} data.
As described in the Introduction, this is done by introducing {and varying} two additional parameters, $A^{\rm TTTEEE}_{\rm L}$ and $A^{\phi\phi}_{\rm L}$, as in Refs.~\cite{Simard:2017xtw,Wu:2019hek,Murgia:2020ryi,Abellan:2021bpx}.
The results are shown in Fig.~\ref{Fig:Pseudo_Alens} and in the second column of Tab.~\ref{tab:bf1}, from which we note that the lensing anomaly is not relaxed and that the tight constraint on the sterile neutrino mass is not alleviated by the introduction of the two extra parameters. As in the $\Lambda$CDM framework, in this set-up there is no growth tension, due to the anti-correlation of $S_8$ with the lensing parameters.
Moreover, Fig.~\ref{Fig:Pseudo_Alens} clearly shows that the introduction of $A_{\rm{L}}^{\rm{TTTEEE}}$ and $A_{\rm{L}}^{\rm{\phi\phi}}$ introduces new degeneracies in the parameter space, resulting in higher values for both $n_s$ and $H_0$. Since both of them are positively correlated with $\Delta N_{\rm eff}$, the tight bound on the latter is relaxed, resulting in $\Delta N_{\rm{eff}}=0.37^{+0.17}_{-0.25}$, i.e.~a mild preference for a non-zero value of the number of additional relativistic species. In other words, the presence of the two additional lensing parameters allows us to find higher values along the well-known $H_0$--$\Delta N_{\rm eff}$ degeneracy direction without a significant impact on the CMB observables.

{Let us now recall that within $\Lambda$CDM the CMB lensing anomaly is characterized by two different aspects: (i) the anomalous value of the observed lensing amplitude parameter, discrepant with the model prediction; (ii) the fact that only $A_{\rm{L}}^{\rm{TTTEEE}}$ is anomalous, while $A_{\rm{L}}^{\phi\phi}$ is compatible with 1, implying that the effect on the TT-TE-EE spectra that seems to be due to an extra-source of gravitational lensing cannot be attributed to actual gravitational lensing. From Fig.~\ref{Fig:Pseudo_Alens} one can notice that in the pseudoscalar model both the lensing parameters depart from 1, indicating that in such a scenario the condition (ii) is not verified. Namely, the extra-smoothing of the peaks is still present but might be attributed to actual lensing. 
Nonetheless, this result should be taken with great care, given that $A_{\rm{L}}^{\phi\phi}$ is again fully compatible with 1 as soon as one adds to the analysis complementary low-redshift data from BAO and SnIa, as we will show in the next Subsection.}

\subsection{Impact of additional low-redshift data}\label{sec:lowred}

\begin{figure}[ht]
 \centering
   \includegraphics[scale=0.4]{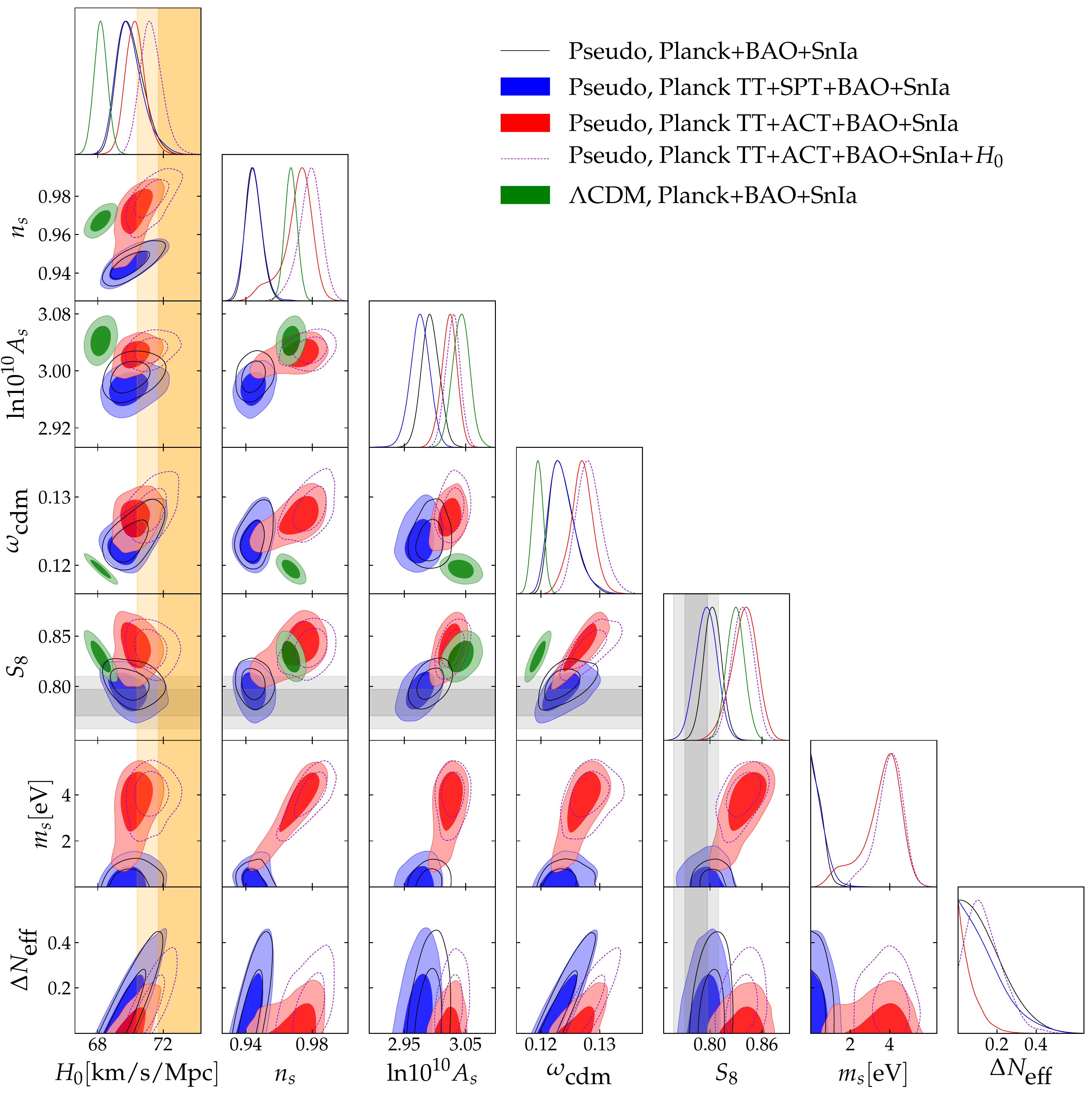} %
  \caption{Posterior distributions of the cosmological parameters in the $\Lambda$CDM and pseudoscalar (Pseudo) model for different data-sets. The orange and gray bands represent the direct measurements (1 $\sigma$ and 2 $\sigma$ confidence regions) of $H_0$ and $S_8$, from cosmic distance ladder~\cite{Riess:2020fzl} and weak lensing observations~\cite{Gatti:2021uwl}, respectively.
  }
  \label{Fig:pseudo_LCDM_Planck_SPT_ACT_BAO}
\end{figure}

 \begin{figure}[t]
 \centering
   \includegraphics[scale=0.4]{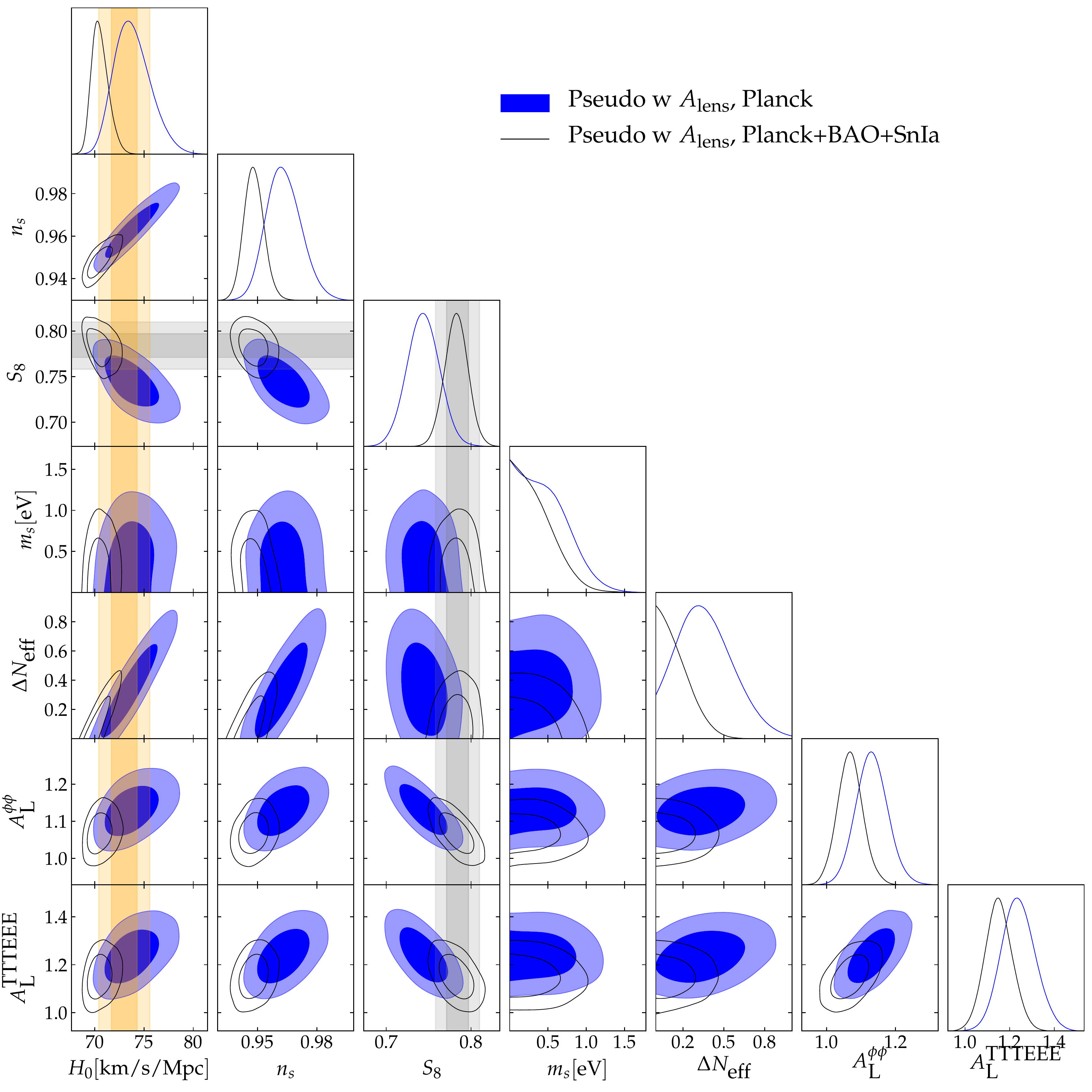}  
  \caption{\looseness=-1 Posterior distributions of the cosmological parameters in the pseudoscalar (Pseudo) model tested with \emph{Planck} data marginalizing over the gravitational lensing information, with and without the addition of BAO and SnIa. The orange and gray bands represent the direct measurements (1 $\sigma$ and 2 $\sigma$ confidence regions) of $H_0$ and $S_8$, from cosmic distance ladder~\cite{Riess:2020fzl} and weak lensing observations~\cite{Gatti:2021uwl}, respectively. }
  \label{Fig:Pseudo_Alens_BAO}
\end{figure}

\begin{table}[hb]
\centering
\scalebox{0.73}{
  \begin{tabular}{|l||c|c|c||c|c|c|c|}
    \hline

& \multicolumn{5}{c|}{{Pseudo vs CMB + BAO + SnIa}}\\ \hline\hline
Parameter & \emph{Planck} & \emph{Planck} w. $A_\textrm{lens}$ & \emph{Planck} TT+SPT  & \emph{Planck} TT+ACT & +$H_0$ \\
\hline \hline

100 $\omega_b$ & $2.259_{-0.019}^{+0.022}$  & $2.276_{-0.020}^{+0.023}$ & $2.269_{-0.023}^{+0.032}$ & $2.209_{-0.021}^{+0.018}$ & $2.225_{-0.020}^{+0.018}$
\\
$\omega_{\rm{cdm}}$ &  $0.1237_{-0.0025}^{+0.0015}$ & $0.1225_{-0.0026}^{+0.0016}$ & $0.1236_{-0.0027}^{+0.0016}$ & $0.1269_{-0.0018}^{+0.0020}$ & $0.1282_{-0.0025}^{+0.0019}$
\\
100 $\theta_s$ & $1.04520_{-0.00037}^{+0.00037}$ & $1.04540_{-0.00037}^{+0.00036}$ & $1.04490_{-0.00047}^{+0.00043}$ & $1.04660_{-0.00041}^{+0.00051}$ & $1.04680_{-0.00042}^{+0.00043}$
\\
$\textrm{ln}10^{10}A_s$ &  $2.992_{-0.015}^{+0.014}$ & $2.971_{-0.016}^{+0.018}$ & $2.976_{-0.016}^{+0.016}$ & $3.023_{-0.012}^{+0.013}$ & $3.025_{-0.012}^{+0.011}$
\\
$n_s$ &  $0.9444_{-0.0053}^{+0.0043}$  & $0.9482_{-0.0053}^{+0.0047}$ & $0.9448_{-0.0055}^{+0.0049}$ & $0.971_{-0.006}^{+0.011}$ & $0.9786_{-0.0055}^{+0.0076}$
\\
$\tau_{\textrm{reio}}$ & $0.0556_{-0.0072}^{+0.0069}$ & $0.0480_{-0.0074}^{+0.0085}$ & $0.0492_{-0.0074}^{+0.0080}$ & $0.0520_{-0.0058}^{+0.0056}$ &  $0.0532_{-0.0062}^{+0.0052}$
\\
$m_s$ [eV] & $<1.0$  & $<0.8$ & $<1.3$  & $3.6_{-0.5}^{+1.1}$ & $3.992_{-0.55}^{+0.77}$
\\
$\Delta N_{\textrm{eff}}$ & $<0.36$  & $<0.38$ & $<0.37$  & $<0.18$ & $0.14_{-0.13}^{+0.05}$
\\
$A^{\phi\phi}_{\rm{L}}$ & & $1.069_{-0.037}^{+0.036}$ & & & 
\\
$A^{\rm{TTTEEE}}_{\rm{L}}$ & & $1.151_{-0.064}^{+0.060}$ & & & 
\\
\hline \hline
$H_0$ [km/s/Mpc] & $70.0_{-0.9}^{+0.7}$  & $70.6_{-1.0}^{+0.7}$ & $69.7_{-1.0}^{+0.7}$ & $70.3_{-0.7}^{+0.6}$ & $71.3_{-0.8}^{+0.6}$
\\
$S_8$ & $0.803_{-0.011}^{+0.010}$ & $0.783_{-0.014}^{+0.014}$ & $0.796_{-0.013}^{+0.013}$ & $0.839_{-0.013}^{+0.017}$ & $0.838_{-0.013}^{+0.014}$
\\
\hline
\end{tabular}}
\caption{The mean $\pm 1~\sigma$ error ($2~\sigma$ in the case of
upper bounds) of the cosmological parameters from CMB experiments in the pseudoscalar model.
}
\label{tab:bf2}
\end{table}

 \begin{figure}[h!]
 \centering
   \includegraphics[scale=0.4]{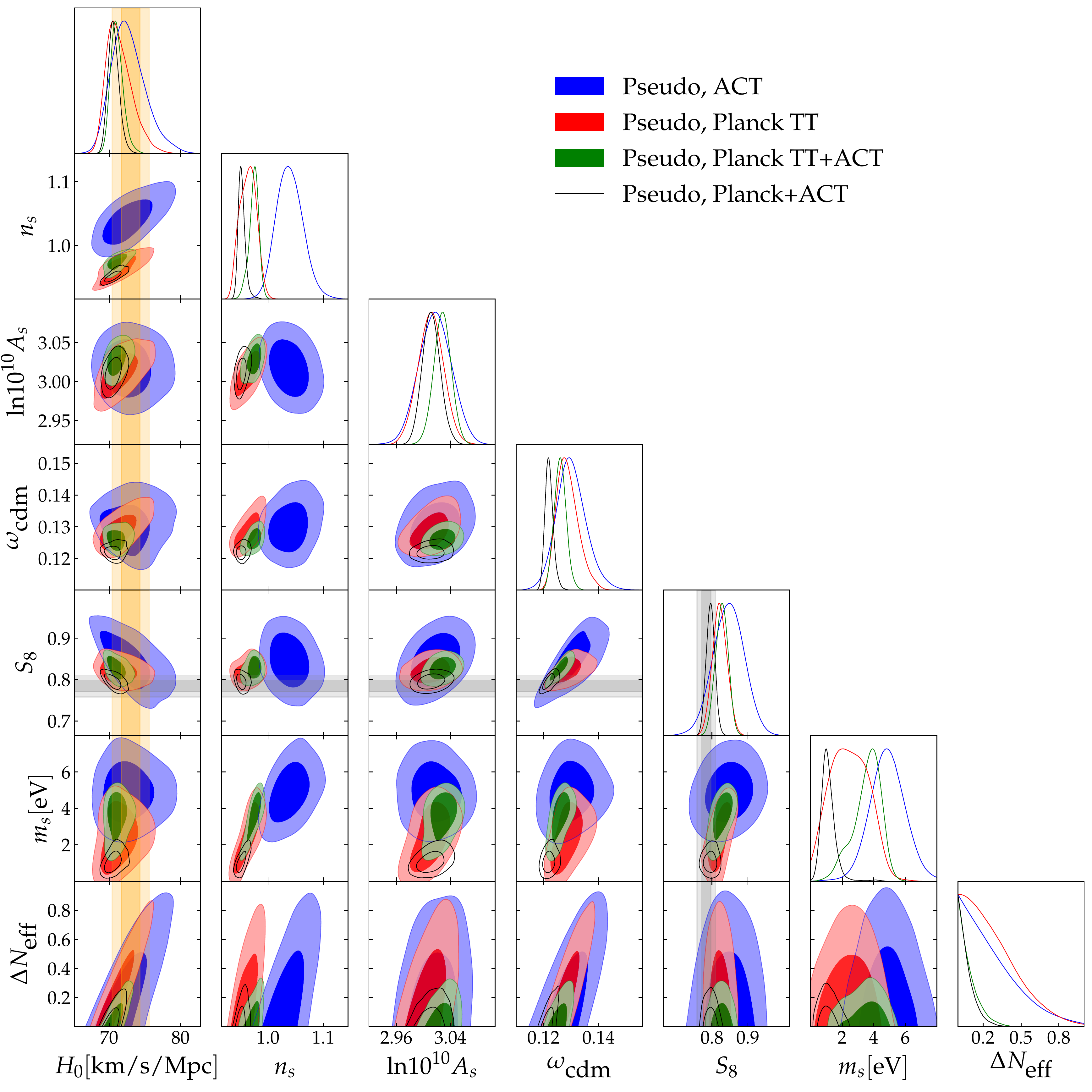}  
  \caption{Posterior distributions of the cosmological parameters in the pseudoscalar (Pseudo) model for different data-sets. The orange and gray bands represent the direct measurements (1 $\sigma$ and 2 $\sigma$ confidence regions) of $H_0$ and $S_8$, from cosmic distance ladder~\cite{Riess:2020fzl} and weak lensing observations~\cite{Gatti:2021uwl}, respectively.}
  \label{Fig:act_new}
\end{figure}

We now examine the impact of adding low-redshift data from BAO and SnIa to the various CMB set-ups presented in the previous Subsections. In Fig.~\ref{Fig:pseudo_LCDM_Planck_SPT_ACT_BAO} we show how the limits on the pseudoscalar model displayed in Fig.~\ref{Fig:pseudo_LCDM_Planck_SPT_ACT} are affected by the addition of BAO and SnIa data, whereas in Fig.~\ref{Fig:Pseudo_Alens_BAO} we show the impact of such low-redshift probes on the predictions within the pseudoscalar scenario when marginalizing over the lensing parameters as we did in Subsection~\ref{sec:Planck+Alens}. The corresponding mean values and $\pm 1 \sigma$~C.L.\ are reported in Tab.~\ref{tab:bf2}.
The addition of BAO+SnIa data does not significantly shift the posteriors of the cosmological parameters. Concerning the \textit{Planck} analysis, our results are in {excellent} agreement with those from Ref.~\cite{Archidiacono:2020yey}, with minor differences due to a slightly different choice for the BAO data-set. In the \textit{Planck} TT+SPT analysis, the additional constraining power coming from low-redshift data further tightens the limits on the $m_s$ and $\Delta N_{\rm eff}$ parameters.
In the \textit{Planck} TT+ACT case, the preference for a non-zero value for $m_s$ is not affected by the addition of low-redshift data, although the global fit is slightly degraded ($\Delta\chi^2\simeq6$ instead of $\Delta\chi^2\simeq3$, compared to $\Lambda$CDM) with respect to the case  without BAO+SnIa.
Both in Fig.~\ref{Fig:pseudo_LCDM_Planck_SPT_ACT_BAO} and Tab.~\ref{tab:bf2}, we also report the results of a \textit{Planck} TT+ACT+BAO+SnIa analysis carried out adopting an effective calibration prior on the absolute magnitude ($M_B$) of SnIa~\cite{Camarena:2019moy,Camarena:2021jlr} that corresponds to the direct determination of $H_0=73.04^{+1.04}_{-1.04}$ km/s/Mpc from the SH0ES Collaboration~\cite{Riess:2021jrx}\footnote{\textbf{ \url{https://github.com/valerio-marra/CalPriorSNIa}.}}. 
By comparing the full red and empty purple contours in Fig.~\ref{Fig:pseudo_LCDM_Planck_SPT_ACT_BAO}, we note that the preference for a larger value of $n_s$ is cleaner, with a clear peak around $n_s=0.9786_{-0.0055}^{+0.0076}$. The tails at low values of both $n_s$ and $m_s$ disappear, and there is now a mild hint for a non-zero $\Delta N_{\rm eff} = 0.14_{-0.13}^{+0.05} $, due to the well-known positive correlation of the latter with the Hubble parameter.
The corresponding value of $H_0 = 71.3^{+0.6}_{-0.8}$ km/s/Mpc is in good agreement with its local measurement. We obtain indeed a total $\Delta\chi^2\simeq-10$ with respect to the $\Lambda$CDM model and, according to the Akaike Information Criterion~\cite{Akaike1974ANL} (AIC), this is the only analysis discussed so far where the pseudoscalar model is preferred over $\Lambda$CDM (see Tab.~\ref{tab:chi2_Pseudo_CMB_BAO_SnIa}).
As anticipated, in Fig.~\ref{Fig:Pseudo_Alens_BAO} we show that, although in the \textit{Planck}-only analysis of the pseudoscalar model both the lensing parameters are anomalous (see Sec.~\ref{sec:Planck+Alens}), the addition of BAO+SnIa data shifts the posterior distributions of $A^{\phi\phi}_{\rm{L}}$ towards the $\Lambda$CDM prediction, such that $A^{\phi\phi}_{\rm{L}}$ is again fully compatible with 1.
Moreover, the addition of low-redshift probes has the effect of reducing the parameter space opened by the introduction of free $A^{\rm TTTEEE}_{\rm{L}}$ and $A^{\phi\phi}_{{\rm{L}}}$ in the CMB-only analysis, such that the prediction of $n_s$, $H_0$ and $\Delta N_{\rm{eff}}$ are now close to the one obtained in the $Planck$+BAO+SnIa case (see also Tab.~\ref{tab:bf2}).
The price to pay for the changes in the parameter values induced to accommodate BAO+SnIa data, is a slight degradation of the fit to \textit{Planck} high-$\ell$ TT data ($\Delta\chi^2\simeq10$ instead of $\Delta\chi^2\simeq7$, compared to $\Lambda$CDM).

\section{\label{sec:act2}Planck vs ACT: a deeper look}

 \begin{figure}[h!]
 \centering
 \subfigure{\includegraphics[scale=0.465]{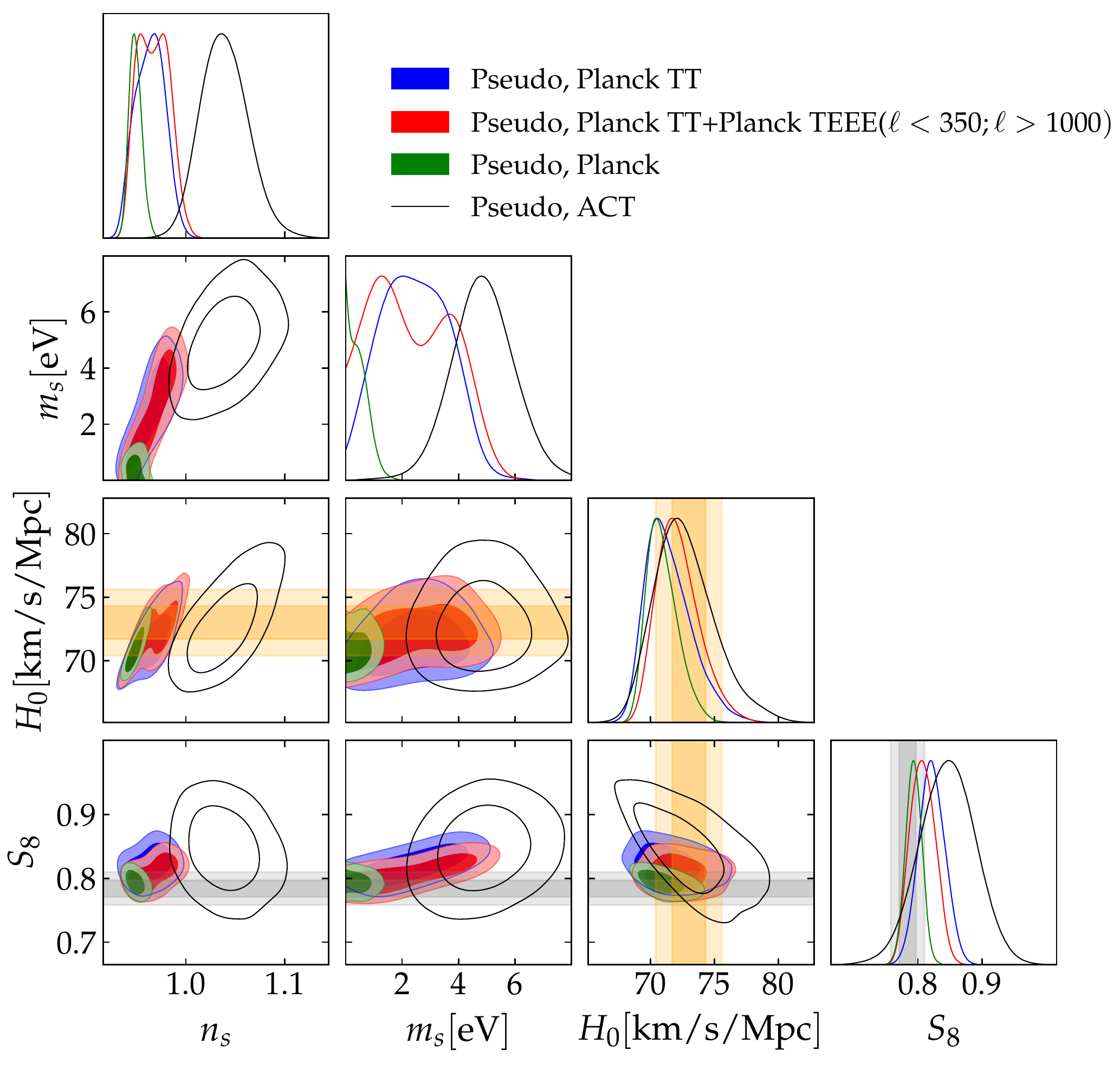}}
 \subfigure{\includegraphics[scale=0.465]{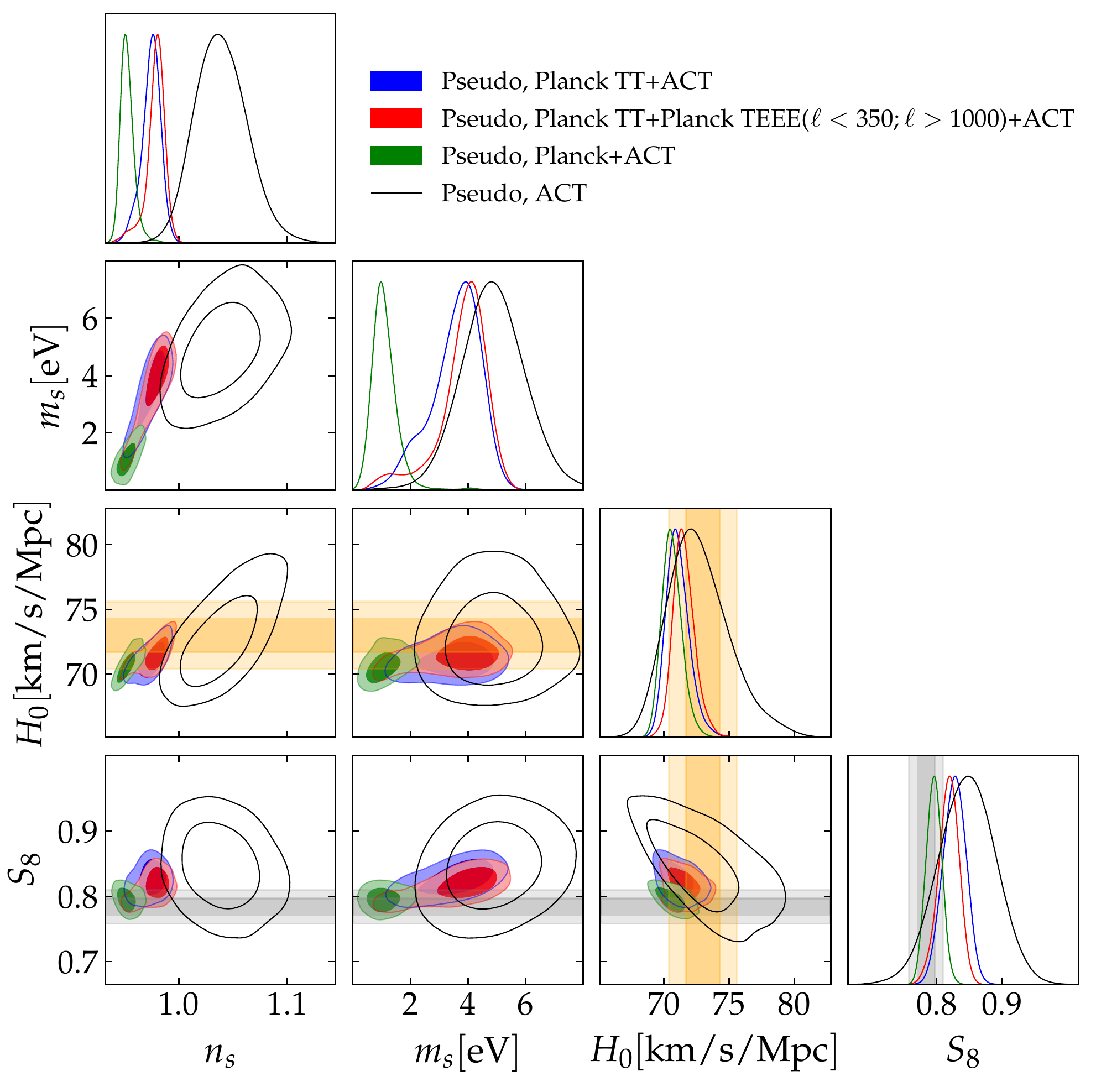}}
 \caption{Posterior distributions of the cosmological parameters in the pseudoscalar (Pseudo) model for different subsets of \emph{Planck} TE-EE data with and without the inclusion of ACT. The orange and gray bands represent the direct measurements (1 $\sigma$ and 2 $\sigma$ confidence regions) of $H_0$ and $S_8$, from cosmic distance ladder~\cite{Riess:2020fzl} and weak lensing observations~\cite{Gatti:2021uwl}, respectively.
 }

\label{fig:Planckcut}
\end{figure}

As explained in the previous Section, our results explicitly demonstrate that the approximate consistency of cosmological parameters inferred from \emph{Planck} and ACT in the $\Lambda$CDM framework~\cite{Aiola:2020azj} does not necessarily imply their consistency in more complex models. In fact, the apparent (mild) tension between \emph{Planck} and ACT data in the EE and TE spectra at around $ \ell \sim 500$ could have non-trivial implications in extended models compared to $\Lambda$CDM, as it was also pointed out very recently within scenarios featuring dark energy at early times~\cite{Poulin:2021bjr,Hill:2021yec}. In this Section, we further investigate the source of the tension between \textit{Planck} and ACT polarization data, that in the model under study is directly responsible for the preference for a non-zero sterile neutrino mass in the joint \emph{Planck} TT+ACT analysis. To that aim, we performed a set of additional MCMC runs, summarized in Fig.~\ref{Fig:act_new} and Tab.~\ref{tab:bf3}.
We tested the pseudoscalar model against ACT data alone, \emph{Planck} low-$\ell$ + high-$\ell$ TT data as in Ref.~\cite{Archidiacono:2020yey}, and the combination of ACT data with the full data-set from \emph{Planck}. Hereafter, we label them as ACT, \emph{Planck} TT and \emph{Planck}+ACT, respectively.
Let us note that the pseudoscalar model provides a better fit with respect to $\Lambda$CDM in the ACT-only analysis ($\Delta\chi^2 \simeq -5$), and appears also to be mildly favoured according to the AIC, although this preference disappears as soon as BAO+SnIa are added to the analysis (see Tabs.~\ref{tab:chi2_Pseudo_CMB} and \ref{tab:chi2_Pseudo_CMB_BAO_SnIa}).
In Sec.~\ref{sec:act1} we pointed out that the preference for a non-zero $m_s$ is driven by the preference for higher values of $n_s$ from ACT data, compared to \emph{Planck} and SPT, as already {discussed} by the ACT Collaboration in the $\Lambda$CDM context~\cite{Aiola:2020azj}.
Indeed, the relatively low constraining power from ACT on large angular scales results in a strong anti-correlation between $\omega_b$ and $n_s$. Hence, a lower value of the baryon density, which damps the small-scale power spectrum, can be partially compensated by a higher value of the spectral index which tilts the spectrum to restore the small-scale power~\cite{Aiola:2020azj}.
Within that degeneracy direction, ACT data favour a lower $\omega_b$ and a higher $n_s$, with respect to \emph{Planck}.
As one can see from Fig.~\ref{Fig:act_new} and the first two columns of Tab.~\ref{tab:bf3}, ACT data alone prefer a value of $n_s$ greater than one {even in the pseudoscalar scenario}.
In light of these considerations, our choice to {consider} the joint \emph{Planck} TT+ACT as our reference {data combination}, as we did in Sec.~\ref{sec:act1}, appears the most natural one, since in this data-set the posterior of $n_s$ is shifted towards values lower than one. Strikingly, this shift only marginally affects the posterior of $m_s$, so that a non-zero mass is still preferred.
Let us also notice that neither ACT nor \emph{Planck} TT data alone put very tight limits on $\Delta N_{\rm{eff}}$, whereas their joint analysis makes the constraint on this parameter very stringent, due to the broken degeneracy in the $\Delta N_{\rm{eff}}-n_s$ plane.
{From Fig.~\ref{Fig:act_new}} we can also note that the posterior distributions of the \emph{Planck} TT analysis agree very well with the results from Ref.~\cite{Archidiacono:2020yey}, and are always in good agreement with those from ACT-only, ensuring the statistical consistency of the joint \emph{Planck} TT+ACT analysis. As we also stressed in Sec.~\ref{sec:bf_cosmo}, the predictions of our reference analysis are indeed very similar to those from \emph{Planck} TT data only, besides featuring narrower posteriors.

Let us now examine the \emph{Planck}+ACT results. As expected, the posteriors of $m_s$ and $n_s$ sit midway between the ones obtained from individually considering ACT or \emph{Planck}, predicting a lower, but still non-zero value for the sterile mass,~i.e.~$m_s=1.1^{+0.3}_{-0.5}$~eV (68\%~C.L.), and a lower $S_8$ value.
There is of course a strong degradation of the global fit with respect to $\Lambda$CDM ($\Delta \chi^2 \simeq 16$)
driven indeed by a poor fit to both ACT and \emph{Planck} data.
However, these latter results must be considered only as a proof-of-principle, given that such a data combination is not fully statistically consistent.
As one can notice by comparing the first column of Tab.~\ref{tab:bf1} with the first column of Tab.~\ref{tab:bf3}, the two data-sets predict values for $n_s$ and $m_s$ which are in disagreement at $\gtrsim 3\sigma$.

In order to identify which subset of \textit{Planck} TE-EE data drives the tension with ACT, making the combination of the full \emph{Planck} data-set with ACT statistically inconsistent, we performed a further set of analyses where we imposed different cuts in the multipole range of \emph{Planck} TE-EE data, as shown in Fig.~\ref{fig:Planckcut} (see also App.~\ref{App:C}).
The results reported in Fig.~\ref{fig:Planckcut} clearly illustrate that it is possible to find a subset of \textit{Planck} polarization data that, when combined with \emph{Planck} TT, is in statistical agreement with ACT. 
From the upper panel of Fig.~\ref{fig:Planckcut} we note indeed that when we ``restrict'' \emph{Planck} TE-EE data by excluding multipoles between $350 < \ell < 1000$, we obtain parameter bounds in agreement at approximately $1~\sigma$ with ACT predictions. 
Hence, the tight constraint on $n_s$, inconsistent with the values favoured by ACT, is driven by \emph{Planck} TE-EE data in that multipole range. 
In the lower panel of Fig.~\ref{fig:Planckcut} we report the results of the same ``restricted'' \emph{Planck} analysis, but in combination with ACT.

Both analyses are fully {statistically} consistent, and the corresponding contours in excellent agreement with those from \emph{Planck} TT+ACT.
Therefore, we can conclude that the possibility of a non-zero sterile neutrino mass is primarily excluded by \emph{Planck} polarization data in the multipole range between $350\lesssim\ell\lesssim1000$, also in accordance from what one can intuitively guess from Fig.~\ref{fig: Residual}. The range of TE-EE multipoles where ACT and \emph{Planck} residuals significantly differ from each other is indeed $350\lesssim\ell\lesssim1000$, and on such intermediate multipoles the \emph{Planck} TT+ACT best-fit pseudoscalar model features the enhanced oscillation pattern with respect to $\Lambda$CDM that we discussed in detail in Sec.~\ref{sec:bf_cosmo}.
Intriguingly, that interval of multipoles plays a crucial role also in the detection of a non-zero fraction of early dark energy when ACT data are taken into account~\cite{Hill:2021yec,Poulin:2021bjr}.
Future work including higher precision data from current and upcoming CMB ground-based experiments will be crucial to test these results, and to understand whether the mild tension between \emph{Planck} and ACT is due to some unaccounted systematics or a statistical fluke.

\begin{table}[t]
\centering
\scalebox{0.72}{
  \begin{tabular}{|l||c|c||c|c|c|c|c|}
    \hline

& \multicolumn{4}{c|}{{Pseudoscalar model}}\\ \hline\hline
Parameter & ACT & +BAO+SnIa & \emph{Planck}+ACT & +BAO+SnIa\\
\hline \hline

100 $\omega_b$ & $2.185_{-0.037}^{+0.035}$ & $2.175_{-0.035}^{+0.032}$ & $2.239_{-0.020}^{+0.017}$ & $2.238_{-0.019}^{+0.018}$
\\
$\omega_{\rm{cdm}}$ & $0.1301_{-0.0059}^{+0.0049}$ & $0.1311_{-0.0042}^{+0.0024}$ & $0.1220_{-0.0018}^{+0.0013}$ & $0.1226_{-0.0015}^{+0.0011}$
\\
100 $\theta_s$ & $1.04800_{-0.00081}^{+0.00081}$ & $1.04770_{-0.00074}^{+0.00079}$ & $1.04600_{-0.00046}^{+0.00037}$ & $1.04580_{-0.00038}^{+0.00033}$
\\
$\textrm{ln}10^{10}A_s$ & $3.017_{-0.026}^{+0.024}$ & $3.014_{-0.023}^{+0.025}$ & $3.012_{-0.014}^{+0.013}$ & $3.005_{-0.012}^{+0.012}$
\\
$n_s$ & $1.039_{-0.026}^{+0.022}$ & $1.022_{-0.021}^{+0.019}$ & $0.9527_{-0.0077}^{+0.0047}$ & $0.9475_{-0.0050}^{+0.0038}$
\\
$\tau_{\textrm{reio}}$  & $0.06_{-0.01}^{+0.01}$ & $0.055_{-0.009}^{+0.010}$ & $0.0582_{-0.0065}^{+0.0059}$ & $0.0558_{-0.0060}^{+0.0056}$
\\
$m_s$ [eV] & $4.9_{-1.2}^{+1.1}$ & $4.6_{-1.1}^{+1.2}$ & $1.1_{-0.5}^{+0.3}$ & $0.9_{-0.4}^{+0.3}$
\\
$\Delta N_{\textrm{eff}}$  & $<0.75$ & $<0.47$ & $<0.21$ & $<0.16$
\\
\hline \hline
$H_0$ [km/s/Mpc] & $72.7_{-2.8}^{+2.0}$ & $70.6_{-1.2}^{+0.8}$ & $70.6_{-0.9}^{+0.7}$ & $69.8_{-0.6}^{+0.5}$
\\
$S_8$ & $0.846_{-0.044}^{+0.045}$ & $0.873_{-0.021}^{+0.021}$ & $0.796_{-0.013}^{+0.012}$ & $0.806_{-0.010}^{+0.010}$\\
\hline 

\end{tabular}}
\caption{The mean $\pm 1~\sigma$ error ($2~\sigma$ in the case of
upper bounds) of the cosmological parameters from CMB experiments in the pseudoscalar model.
}
\label{tab:bf3}
\end{table}

\section{Comparison with SBL neutrino oscillation experiments}\label{sec:sbl}

 \begin{figure}[bh]
 \centering
 \includegraphics[scale=0.42]{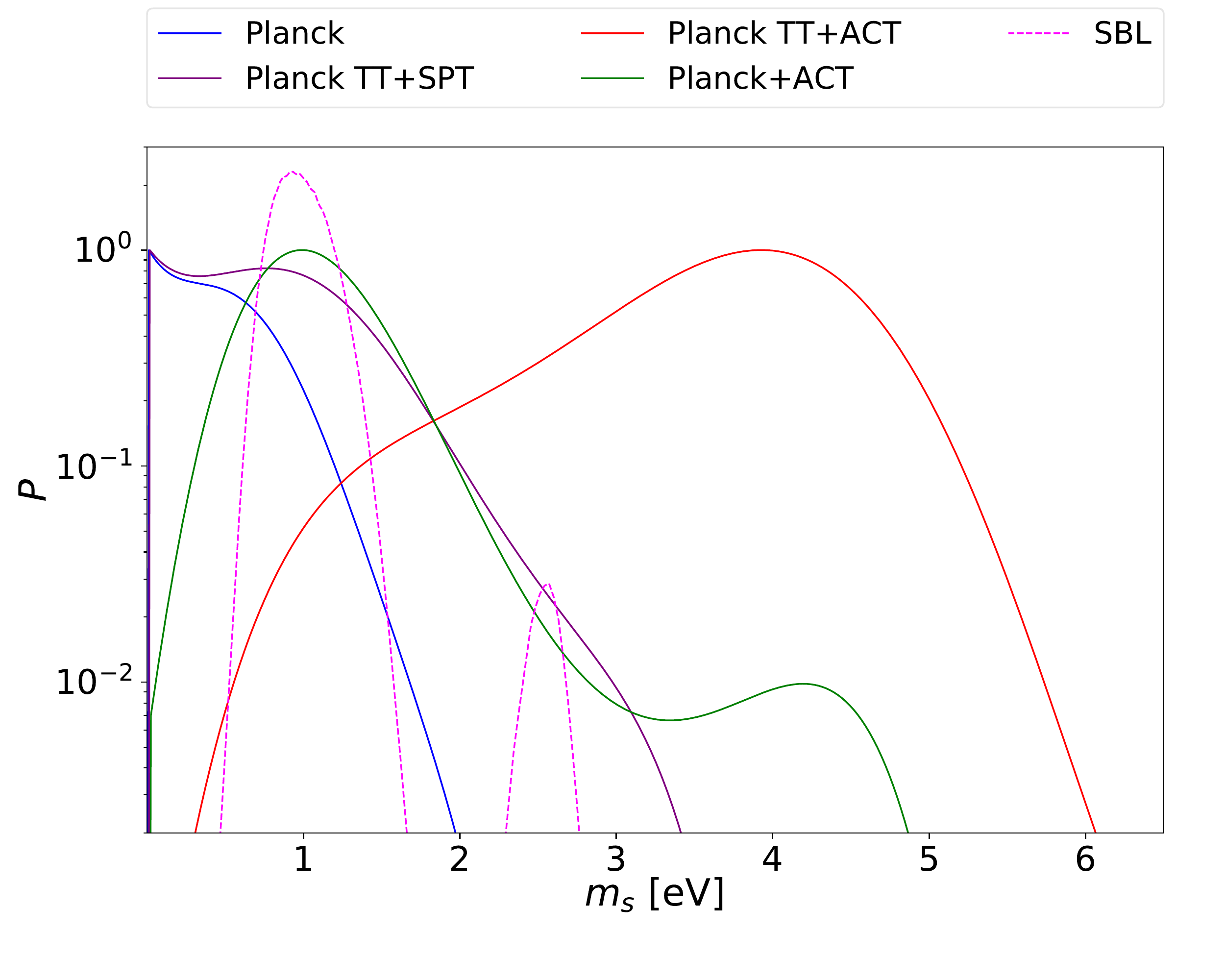}
 \caption{\looseness=-1 Marginalized 1-D posterior distributions for the sterile neutrino mass, in the pseudoscalar model, from the cosmological analyses performed in this work (solid lines) and from SBL data (dashed line). See Sec.~\ref{sec:sbl} for a detailed description of the latter ones.}

\label{fig:sbl}
\end{figure}

As discussed in the Introduction,
SBL neutrino oscillation anomalies
may be explained by the existence of a sterile neutrino at the eV mass scale
(see the recent reviews in Refs.\ \cite{Gariazzo:2015rra,Boser:2019rta,Giunti:2019aiy}).
However,
the tension between appearance and disappearance 
SBL neutrino oscillation data
in the framework of 3+1 active-sterile neutrino mixing~\cite{Dentler:2018sju,Gariazzo:2017fdh,Diaz:2019fwt}
does not allow us to obtain a global SBL fit.
Moreover,
the reactor antineutrino anomaly has been revised and reduced recently~\cite{Berryman:2020agd,Giunti:2021kab}.
On the other hand,
the Gallium anomaly has been reinforced by the recent results of the BEST experiment~\cite{Barinov:2021asz},
but it is in tension with the reactor data~\cite{Giunti:2021kab},
with
the negative results of the
DANSS~\cite{Danilov:2020ucs},
PROSPECT~\cite{PROSPECT:2020sxr}, and
STEREO~\cite{STEREO:2019ztb}
experiments~\cite{Barinov:2021mjj},
and with the solar bound~\cite{Goldhagen:2021kxe,Giunti:2021kab,Berryman:2021yan}.
Considering the SBL $\nu_{\mu}\to\nu_{e}$ appearance channel,
the MiniBooNE low-energy anomaly~\cite{MiniBooNE:2018esg,MiniBooNE:2020pnu}
has been recently disfavored by the results of the MicroBooNE experiment~\cite{MicroBooNE:2021jwr,MicroBooNE:2021nxr,MicroBooNE:2021rmx}
(see, however, the caveats in Ref.~\cite{Arguelles:2021meu}
and the possible interpretation of the MicroBooNE data in terms of
SBL $\nu_e$ disappearance in Ref.~\cite{Denton:2021czb}).
Only the LSND excess in the $\bar\nu_{\mu}\to\bar\nu_{e}$ appearance channel~\cite{LSND:2001aii}
has not been excluded by other experiments so far
(it will be soon under investigation in the SBN~\cite{Machado:2019oxb} and JSNS$^2$~\cite{Rott:2018rlw} experiments).
Therefore, we consider the LSND excess~\cite{LSND:2001aii}
taking into account the constraints on the sterile neutrino mass
given by the negative results of the
BNL-E776~\cite{Borodovsky:1992pn},
KARMEN~\cite{KARMEN:2002zcm},
NOMAD~\cite{NOMAD:2003mqg},
ICARUS~\cite{ICARUS:2013cwr}
and
OPERA~\cite{OPERA:2013wvp}
$\nu_{\mu}\to\nu_{e}$ appearance experiments.
The Bayesian posterior given by the combined fit of the data of these experiments
is shown by the dashed line in Fig.~\ref{fig:sbl}.
One can see that there is a main peak around $m_s \approx 1 \, \text{eV}$
and a secondary peak around
$m_s \approx 2.5 \, \text{eV}$.
\looseness=-1
There is a clear compatibility between these values of $m_s$
and the cosmological indications in the pseudoscalar model.

\section{Conclusions}\label{sec:concl}

{In this work we have employed} a whole host of up-to-date cosmological data, both at the background and at the linear perturbation level,
{to test the} cosmological scenario first proposed in Ref.~\cite{Archidiacono:2014nda} and more recently studied in Refs.~\cite{Archidiacono:2015oma,Archidiacono:2016kkh,Archidiacono:2020yey}, where an additional light sterile neutrino species is allowed to self-interact, through a new pseudoscalar mediator. 
This scenario naturally predicts a value of $H_0$ consistent with its local measurements, and it provides a good fit to CMB data without spoiling LSS bounds on the neutrino mass nor worsening the $S_8$ tension with respect to the $\Lambda$CDM model.
{While being slightly favoured by \emph{Planck} temperature only data~\cite{Archidiacono:2016kkh},}
a non-zero sterile neutrino mass -- compatible with the range suggested to explain SBL anomalies -- is however severely constrained by the addition of high-$\ell$ polarization data, {as already pointed out in} Ref.~\cite{Archidiacono:2020yey}.

Given the availability of ground-based CMB maps with better angular resolution than \emph{Planck} on intermediate and high multipoles, our goal was to test the robustness of the tight limits on the sterile neutrino properties from Ref.~\cite{Archidiacono:2020yey}, and at the same time evaluate the impact of the pseudoscalar model on the CMB lensing anomaly, which affects high-$\ell$ \emph{Planck} data.
To this end, we have expanded previous analyses as follows:
(i) showing the impact of marginalizing over the CMB lensing {amplitude} in \emph{Planck} data;
(ii) replacing \emph{Planck} high-$\ell$ polarization data with those from two ground-based CMB surveys,~i.e.~SPT~\cite{Henning:2017nuy} and ACT~\cite{Aiola:2020azj}.
In all our analyses we have found that the model under study is able to resolve the Hubble tension without worsening the growth tension compared to the $\Lambda$CDM model, while we see no impact on the CMB lensing anomaly.
We have also shown that using additional low-redshift data from BAO and uncalibrated SnIa does not qualitatively affect the constraints, besides a general strengthening {of the credible regions}.
Both {when considering a free lensing amplitude} and in the \emph{Planck} TT+SPT analysis, we have demonstrated that the strong bound on the sterile neutrino mass is not relaxed.

Things dramatically change when the model is confronted against ACT data.
This data-set, both alone and in combination with \emph{Planck}, shows indeed a $\gtrsim 3~\sigma$ preference for a non-zero sterile neutrino mass in the eV range.
In particular, in the \emph{Planck} TT+ACT analysis we find $m_s=3.6^{+1.1}_{-0.6}$~eV (68\%~C.L.).
We have shown that this is due to the positive correlation with the high value of the tilt of the primordial power spectrum $n_s$ favoured by ACT -- even within $\Lambda$CDM -- compared to \emph{Planck} and SPT.
The pseudoscalar model appears indeed to provide a better fit to ACT data with respect to $\Lambda$CDM ($\Delta\chi^2 \simeq -5$).
Interestingly, we have found that the values of $n_s$ and $m_s$ inferred from the entire \emph{Planck} data-set differ at \textbf{$\gtrsim 3\sigma$} from those predicted by ACT alone, making a full \emph{Planck}+ACT analysis statistically inconsistent.
We have thus carried out a further set of studies aimed {at identifying} which subset of \emph{Planck} polarization data is responsible for this tension.
We have explicitly shown that the values of $n_s$ and $m_s$ favoured by multipoles higher (lower) than 1000 (350) from \emph{Planck} TE-EE data are still consistent with ACT predictions, meaning that the severe constraint on the sterile neutrino mass is driven by \emph{Planck} TE-EE data in the multipole range $350\lesssim\ell\lesssim1000$.
The slight discrepancy between \emph{Planck} and ACT on intermediate multipoles does not have a major impact within the $\Lambda$CDM model~\cite{Aiola:2020azj}. It can however lead to highly non-trivial results not only within the self-interacting sterile neutrino scenario studied here, but also in other alternative {models that could resolve} to the Hubble tension, such as early dark energy scenarios~\cite{Hill:2021yec,Poulin:2021bjr,Smith:2022hwi,Jiang:2022uyg}. 
It will thus be crucial to establish whether the discrepancy is due to yet unaccounted systematics or a statistical fluke either in \emph{Planck} or in ACT data.
Our results highlight the {importance} of testing non-trivial extensions of the $\Lambda$CDM model against upcoming, more accurate measurements of the CMB spectra from future surveys, such as CMB-S4~\cite{CMB-S4:2016ple} or the Simons Observatory~\cite{SimonsObservatory:2018koc}, as well as new data releases from both ACT and SPT~\cite{SPT-3G:2021eoc}.

\appendix

\acknowledgments
The authors are thankful to Vivian Poulin and Guillermo F. Abell\'an for many useful comments and discussions. The authors acknowledge the use of computational resources from the Cagliari Unit of INFN and the CNRS/IN2P3 Computing Centre (CC-IN2P3) in Lyon. RM acknowledges the hospitality and support of the Physics Department of the University of Cagliari where part of the project was carried out.
SG acknowledges financial support from the European Union's Horizon 2020 research and innovation programme under the Marie Skłodowska-Curie grant agreement No 754496 (project FELLINI).
The work of CG was supported by the research grant ``The Dark Universe: A Synergic Multimessenger Approach'' number 2017X7X85K under the program PRIN 2017 funded by the Ministero dell'Istruzione, Universit\`a e della Ricerca (MIUR).



\begin{table}[h]
\centering
\scalebox{0.75}{
  \begin{tabular}{|l||c|c|c|c|c|c|c|}
    \hline
& \multicolumn{6}{c|}{{Individual $\chi^2$: $\Lambda$CDM vs CMB}}\\ \hline\hline
& \emph{Planck} & \emph{Planck} w. $A_\textrm{lens}$ & \emph{Planck} TT+SPT & \emph{Planck} TT+ACT & ACT & \emph{Planck}+ACT\\
\hline \hline

{\emph{Planck}}~high$-\ell$ TT,TE,EE & 2346.5 & 2340.1 & - &  - &  - &  2348.8\\
{\emph{Planck}}~high$-\ell$ TT & - & - & 762.0 &  764.6 & - & -  \\
{\emph{Planck}}~low$-\ell$ EE & 395.7 & 395.7 & 395.7 &  396.1 & - & 396.2 \\
{\emph{Planck}}~low$-\ell$ TT & 22.9 & 21.5 & 22.7 &  22.3 & - & 22.2 \\
{\emph{Planck}}~lensing & 9.0 & 8.7 & - &  9.1 & - & 9.1 \\
SPT~high$-\ell$ TE,EE & - & - & 146.8 &  - & - & -  \\
SPT~lensing & - & - & 5.7 &  - & - & -  \\
ACT~DR4~ & - & - & - &  - & 280.1 & -  \\
ACT~DR4~~($\ell_{\rm TT}>1800$) & - & - & - &  239.1 & - & 240.7  \\

\hline \hline
total & 2774.1 & 2766.0 & 1332.9 &  1431.2 & 280.1 & 3017.0  \\
\hline
\end{tabular}}
\caption{Best-fit $\chi^2$ per experiment (and total) from our CMB analyses in the $\Lambda$CDM model.}
\label{tab:chi2_LCDM_CMB}
\end{table}


\begin{table}[tb]
\centering
\scalebox{0.68}{
  \begin{tabular}{|l||c|c|c||c|c||c|c|c}
    \hline

& \multicolumn{7}{c|}{{Individual $\chi^2$: $\Lambda$CDM vs CMB + BAO + SnIa}}\\ \hline\hline
& \emph{Planck} & \emph{Planck} w. $A_\textrm{lens}$ & \emph{Planck} TT+SPT  & \emph{Planck} TT+ACT &+$H_0$ & ACT & \emph{Planck}+ACT\\
\hline \hline

Pantheon SnIa & 1026.9 & 1026.8 & 1026.9 & 1026.9 & 1027.2 & 1027.0 & 1026.9 \\
BAO+FS~BOSS DR12 & 6.2 & 6.3 & 6.1 & 6.2 & 7.3 & 6.5 & 6.4 \\
BAO~BOSS low$-z$ & 1.9 & 2.1 & 2.0 & 1.9 & 2.6 & 1.8 & 1.7 \\
eBOSS\;DR14\;Lyman$-\alpha$ & 4.5 & 4.4 & 4.5 & 4.5 & 4.3 & 4.9 & 4.6 \\
{\emph{Planck}}~high$-\ell$ TT,TE,EE & 2346.5 & 2340.1 & - & - & - & - & 2349.4\\
{\emph{Planck}}~high$-\ell$ TT & - & - & 763.2 & 765.7 & 766.3 & - & - \\
{\emph{Planck}}~low$-\ell$ EE & 395.8 & 395.7 & 395.7 & 395.7 & 396.4 & - & 396.6 \\
{\emph{Planck}}~low$-\ell$ TT & 22.8 & 22.2 & 22.5 & 22.1 & 22.0 & - & 22.2 \\
{\emph{Planck}}~lensing & 8.9 & 9.1 & - & 9.1 & 9.0 & - & 9.3 \\
SPT~high$-\ell$ TE,EE & - & - & 146.6 & - & - & - & -\\
SPT~lensing & - & - & 5.5 & - & - & - & -\\
ACT~DR4~ & - & - & - & - & - & 280.3 & - \\
ACT~DR4~~($\ell_{\rm TT}>1800$) & - & - & - & 238.4 & 238.5 & - & 240.0 \\
$M_B$ Prior (S$H_0$ES 2021) & - & - & - & - & 21.2 & - & -\\

\hline \hline
total & 3813.5 & 3806.7 & 2373.0 & 2470.5 & 2494.8 & 1320.5 & 4057.1 \\
\hline

\end{tabular}}
\caption{Best-fit $\chi^2$ per experiment (and total) from our CMB analyses in combination with BAO and SnIa in the $\Lambda$CDM model.}
\label{tab:chi2_LCDM_CMB_BAO_SnIa}
\end{table}




\begin{table}[tb]
\centering
\scalebox{0.75}{
  \begin{tabular}{|l||c|c|c|c|c|c|c|}
    \hline
& \multicolumn{6}{c|}{{Individual $\chi^2$: Pseudo vs CMB}}\\ \hline\hline
& \emph{Planck} & \emph{Planck} w. $A_\textrm{lens}$ & \emph{Planck} TT+SPT & \emph{Planck} TT+ACT & ACT & \emph{Planck}+ACT\\
\hline \hline

{\emph{Planck}}~high$-\ell$ TT,TE,EE & 2356.3 & 2347.3 & - &  - &  - &  2367.7\\
{\emph{Planck}}~high$-\ell$ TT & - & - & 769.6 &  773.2 &  - &  -\\
{\emph{Planck}}~low$-\ell$ EE & 396.1 & 395.7 & 395.9 &  396.3 &  - &  396.7\\
{\emph{Planck}}~low$-\ell$ TT & 24.9 & 22.0 & 24.0 &  21.2 &  - &  24.0\\
{\emph{Planck}}~lensing & 9.0 & 8.2 & - &  10.8 &  - &  9.0\\
SPT~high$-\ell$ TE,EE & - & - & 149.4 &  - &  - &  -\\
SPT~lensing & - & - & 5.1 &  - &  - &  -\\
ACT~DR4~ & - & - & - &  - &  274.8 &  -\\
ACT~DR4~~($\ell_{\rm TT}>1800$) & - & - & - &  233.2 &  - &  237.0\\

\hline \hline
total & 2786.3 & 2773.2 & 1344.0 &  1434.7 &  274.8 &  3034.4  \\
\hline
total $\Delta \chi^2$ & 12.2 & 7.2 & 11.1 &  3.5 &  -5.3 &  17.4 \\
\hline
\hline
$\Delta \rm{AIC}$ & 16.2 & 11.2 & 15.1 &  7.5 &  -1.3 &  21.4 \\
\hline

\end{tabular}}
\caption{Best-fit $\chi^2$ per experiment (and total) from our CMB analyses in the pseudoscalar model. Per each run, we also report the corresponding $\Delta \chi^2 \equiv \chi^2_{\rm min,pseudo}$-$\chi^2_{\rm min, \Lambda CDM}$. In order to determine if the pseudoscalar model is favoured by the data in the analyses considered, we compute the AIC relative to that of $\Lambda$CDM ($\Delta \rm{AIC}$). Negative values of the latter correspond to a preference for the pseudoscalar model over $\Lambda$CDM.
}
\label{tab:chi2_Pseudo_CMB}
\end{table}


\begin{table}[tb]
\centering
\scalebox{0.68}{
  \begin{tabular}{|l||c|c|c||c|c||c|c|c}
    \hline

& \multicolumn{7}{c|}{{Individual $\chi^2$: Pseudo vs CMB + BAO + SnIa}}\\ \hline\hline
& \emph{Planck} & \emph{Planck} w. $A_\textrm{lens}$ & \emph{Planck} TT+SPT  & \emph{Planck} TT+ACT &+$H_0$ & ACT & \emph{Planck}+ACT\\
\hline \hline

Pantheon SnIa & 1026.9 & 1026.9 & 1026.9 & 1026.9 &  1027.0 &  1027.2 &  1026.9 \\
BAO+FS~BOSS DR12 & 5.7 & 6.9 & 5.9 &  7.1 &  8.4 &  7.7 &  6.0 \\
BAO~BOSS low$-z$ & 1.8 & 2.7 & 2.1 &  2.2 &  2.8 &  1.7 &  2.2 \\
eBOSS\;DR14\;Lyman$-\alpha$ & 4.8 & 4.4 & 4.6 &  5.0 &  4.8 &  5.7 &  4.8 \\
{\emph{Planck}}~high$-\ell$ TT,TE,EE & 2356.7 & 2350.6 & - &  - &  - &  - &  2365.4 \\
{\emph{Planck}}~high$-\ell$ TT & - & - & 768.5 &  773.8 &  774.2 &  - &  - \\
{\emph{Planck}}~low$-\ell$ EE & 396.7 & 395.7 & 395.7 &  395.8 &  395.8 &  - &  395.7\\
{\emph{Planck}}~low$-\ell$ TT & 25.5 & 25.2 & 24.3 & 21.3 &  21.0 &  - &  25.1 \\
{\emph{Planck}}~lensing & 8.6 & 8.2 & - &  10.9 &  10.9 &  - &  8.6\\
SPT~high$-\ell$ TE,EE & - & - & 147.8 &  - &  - &  - &  -\\
SPT~lensing & - & - & 5.2 &  - &  - &  - &  - \\
ACT~DR4~ & - & - & - &  - &  - &  275.2 &  -\\
ACT~DR4~~($\ell_{\rm TT}>1800$) & - & - & - &  233.2 &  233.7 &  - &  239.3\\
$M_B$ Prior (S$H_0$ES 2021) & - & - & - &  - &  6.6 &  - &  -\\

\hline \hline
total & 3826.7 & 3820.6 & 2381.0 &  2476.2 &  2485.2 &  1317.5 &  4074.0\\
\hline
total $\Delta \chi^2$ & 13.2 & 13.9 & 8.0 &  5.7 &  -9.6 &  -3.0 & 16.9 \\
\hline
\hline
$\Delta \rm{AIC}$ & 17.2 & 17.9 & 12.0 &  9.7 &  -5.6 &  1.0 & 20.9 \\
\hline

\end{tabular}}
\caption{Best-fit $\chi^2$ per experiment (and total) from our CMB analyses in combination with BAO and SnIa in the pseudoscalar model. Per each run, we also report the corresponding $\Delta \chi^2 \equiv \chi^2_{\rm min,pseudo}$-$\chi^2_{\rm min, \Lambda CDM}$. In order to determine if the pseudoscalar model is favoured by the data in the analyses considered, we compute the AIC relative to that of $\Lambda$CDM ($\Delta \rm{AIC}$). Positive values prefer the $\Lambda$CDM model.
}
\label{tab:chi2_Pseudo_CMB_BAO_SnIa}
\end{table}


\section{Best-fit parameter values and $\chi^2$ per experiment}\label{app:bfchi2}

\looseness=-1 In Tabs.~\ref{tab:chi2_LCDM_CMB},  \ref{tab:chi2_LCDM_CMB_BAO_SnIa} and in Tabs.~\ref{tab:chi2_Pseudo_CMB},  \ref{tab:chi2_Pseudo_CMB_BAO_SnIa} we report the $\chi^2_{\rm{min}}$’s obtained respectively for the most significant $\Lambda$CDM and pseudoscalar analyses considered. In order to determine if the pseudoscalar model is favoured by the data in the analyses performed, we computed the AIC~\cite{Akaike1974ANL} relative to that of $\Lambda$CDM as $\Delta \rm{AIC} = 2(\mathcal{N}_{\rm{pseudo}}-\mathcal{N}_{\Lambda\rm{CDM}})+\Delta \chi^2$, where $\Delta \chi^2 \equiv \chi^2_{\rm min,pseudo}$-$\chi^2_{\rm min, \Lambda CDM}$ while $\mathcal{N}_{\rm{pseudo}}$ and $\mathcal{N}_{\Lambda\rm{CDM}}$ are the number of free parameters in the pseudoscalar and $\Lambda$CDM model respectively. The pseudoscalar model is favoured over $\Lambda$CDM when negative values of $\Delta \rm{AIC}$ are found. In Tabs.~\ref{tab:bfapp1} and \ref{tab:bfapp2} we also report the best-fit parameter values.


\begin{table}[tb]
\centering
\scalebox{0.73}{
  \begin{tabular}{|l||c|c|c|c|c|c|c|}
    \hline
& \multicolumn{6}{c|}{{Best-fit Pseudo vs CMB}}\\ \hline\hline
Parameter & \emph{Planck} & \emph{Planck} w. $A_\textrm{lens}$ & \emph{Planck} TT+SPT & \emph{Planck} TT+ACT & ACT & \emph{Planck}+ACT\\
\hline \hline

100 $\omega_b$ & 2.263 & 2.307 & 2.265 & 2.213 & 2.156 & 2.223
\\
$\omega_{\rm{cdm}}$ & 0.1219 & 0.1217 & 0.1226 & 0.1257 & 0.1265 & 0.1217
\\
100 $\theta_s$ & 1.04527 & 1.04527 & 1.04481 & 1.04686 & 1.04823 & 1.04653
\\
$\textrm{ln}10^{10}A_s$ &  2.984 & 2.977 & 2.966 & 3.029 & 3.017 & 3.012
\\
$n_s$ & 0.9435  & 0.9598 & 0.946 & 0.976 & 1.023 & 0.9536
\\
$\tau_{\textrm{reio}}$ & 0.0543 & 0.0519 & 0.0464 & 0.0559 & 0.06 & 0.0569
\\
$m_s$ [eV] &  0.0  & 0.4 & 0.2 & 4.0 & 5.0 & 1.4
\\
$\Delta N_{\textrm{eff}}$ & 0.06  & 0.31 & 0.14 & 0.01 & 0.02 & 0.02
\\
$A^{\phi\phi}_{\rm{L}}$ & & 1.119 & & & &
\\
$A^{\rm{TTTEEE}}_{\rm{L}}$ & & 1.221  & & & & 
\\
\hline \hline
$H_0$ [km/s/Mpc] & 69.3  & 73.1 & 69.9 & 70.4 & 70.5 & 70.4
\\
$S_8$ & 0.804 & 0.775 & 0.788 & 0.836 & 0.853 & 0.798
\\
\hline \hline
$ \Delta \chi^2 $ & 12.3 & 7.2 & 11.1 & 3.4 & -5.3 & 17.4\\
\hline
\end{tabular}}
\caption{Best-fit values of the cosmological parameters from our CMB analyses in the pseudoscalar model. Per each run, we also report the corresponding $\Delta \chi^2 \equiv \chi^2_{\rm min,pseudo}$-$\chi^2_{\rm min, \Lambda CDM}$.
}
\label{tab:bfapp1}
\end{table}

\begin{table}[tb]
\centering
\scalebox{0.73}{
  \begin{tabular}{|l||c|c|c||c|c||c|c|c}
    \hline

& \multicolumn{7}{c|}{{Best-fit Pseudo vs CMB + BAO + SnIa}}\\ \hline\hline
Parameter & \emph{Planck} & \emph{Planck} w. $A_\textrm{lens}$ & \emph{Planck} TT+SPT  & \emph{Planck} TT+ACT &+$H_0$ & ACT & \emph{Planck}+ACT\\
\hline \hline

100 $\omega_b$ & 2.257  & 2.272 & 2.284 & 2.210 & 2.201 & 2.157 & 2.224 
\\
$\omega_{\rm{cdm}}$ &  0.1222 & 0.1207 & 0.1213 & 0.1264 & 0.1271 & 0.1289 & 0.1225 
\\
100 $\theta_s$ & 1.04525 & 1.04555 & 1.04481 & 1.04688 & 1.04678 & 1.04790 & 1.04584 
\\
$\textrm{ln}10^{10}A_s$ &  2.989 & 2.974 & 2.975 & 3.018 & 3.024 & 3.018 & 2.994 
\\
$n_s$ &  0.9416  & 0.9429 & 0.9448 & 0.974 & 0.977 & 1.016 & 0.9453 
\\
$\tau_{\textrm{reio}}$ & 0.0563 & 0.0497 & 0.0517 & 0.0489 & 0.0510 & 0.057 & 0.0501 
\\
$m_s$ [eV] & 0.0 & 0.5 & 0.1 & 4.1 & 3.9 & 4.8 & 0.9 
\\
$\Delta N_{\textrm{eff}}$ & 0.04 & 0.01 & 0.03 & 0.0 & 0.08 & 0.03 & 0.01 
\\
$A^{\phi\phi}_{\rm{L}}$ & & 1.059 & & & & & 
\\
$A^{\rm{TTTEEE}}_{\rm{L}}$ & & 1.151 & & & & & 
\\
\hline \hline
$H_0$ [km/s/Mpc] & 68.9 & 69.6 & 69.2 & 70.1 & 70.8 & 69.6 & 69.2 
\\
$S_8$ & 0.812 & 0.785 & 0.796 & 0.839 & 0.839 &  0.876 & 0.808 
\\
\hline \hline
$ \Delta \chi^2 $ & 13.2 & 14 & 8 & 5.7 & -9.6 & -3.1 & 16.9 \\
\hline
\end{tabular}}
\caption{Best-fit values of the cosmological parameters from our
CMB analyses in combination with BAO and SnIa, in the pseudoscalar model. Per each run, we also report the corresponding $\Delta \chi^2 \equiv \chi^2_{\rm min,Pseudo}$-$\chi^2_{\rm min, \Lambda CDM}$.}
\label{tab:bfapp2}
\end{table}

\section{Additional results with different subsets of Planck polarization data}\label{App:C}

\begin{figure}[tb]
\centering
\subfigure{\includegraphics[scale=0.465]{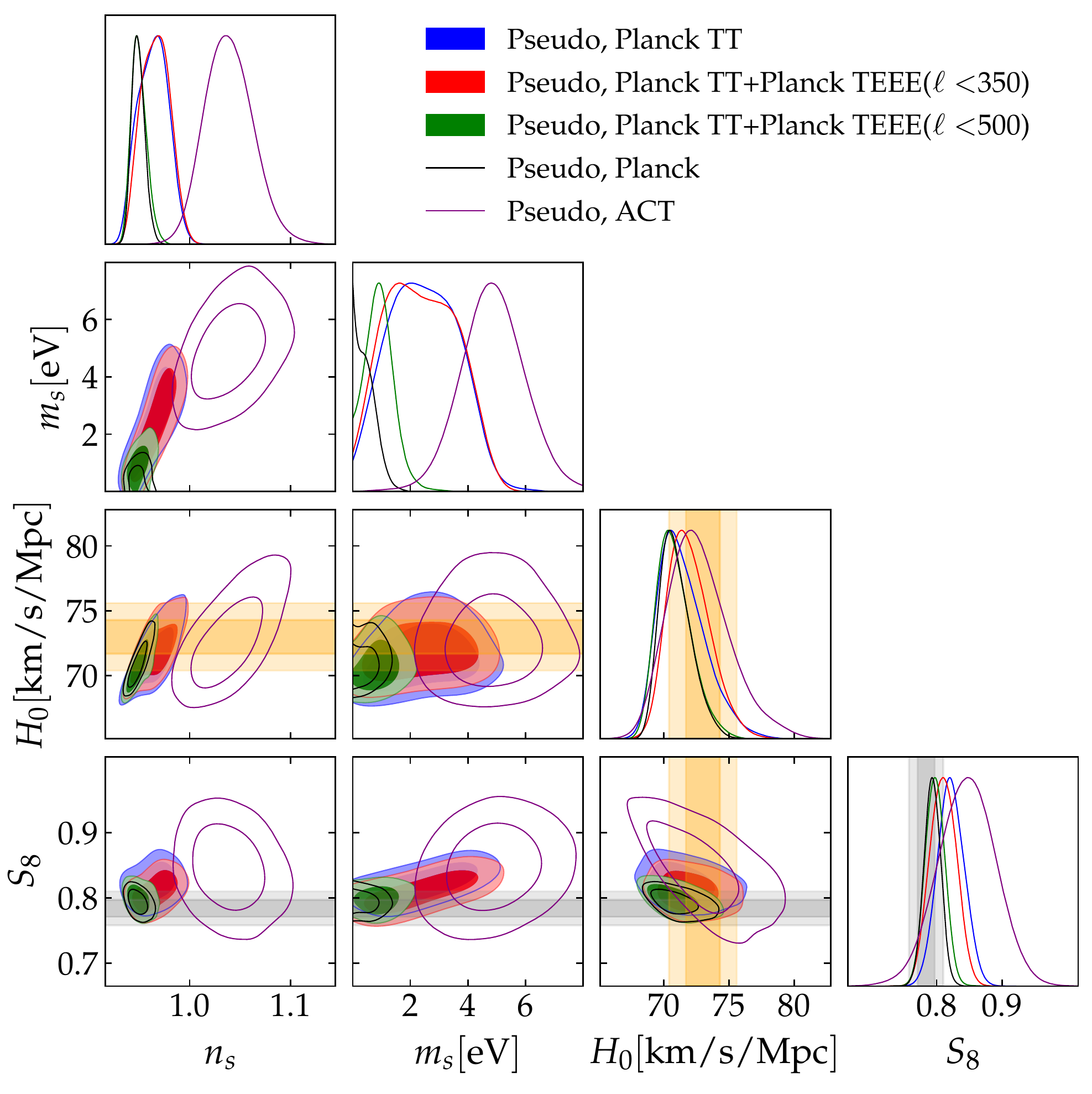}}
\subfigure{\includegraphics[scale=0.465]{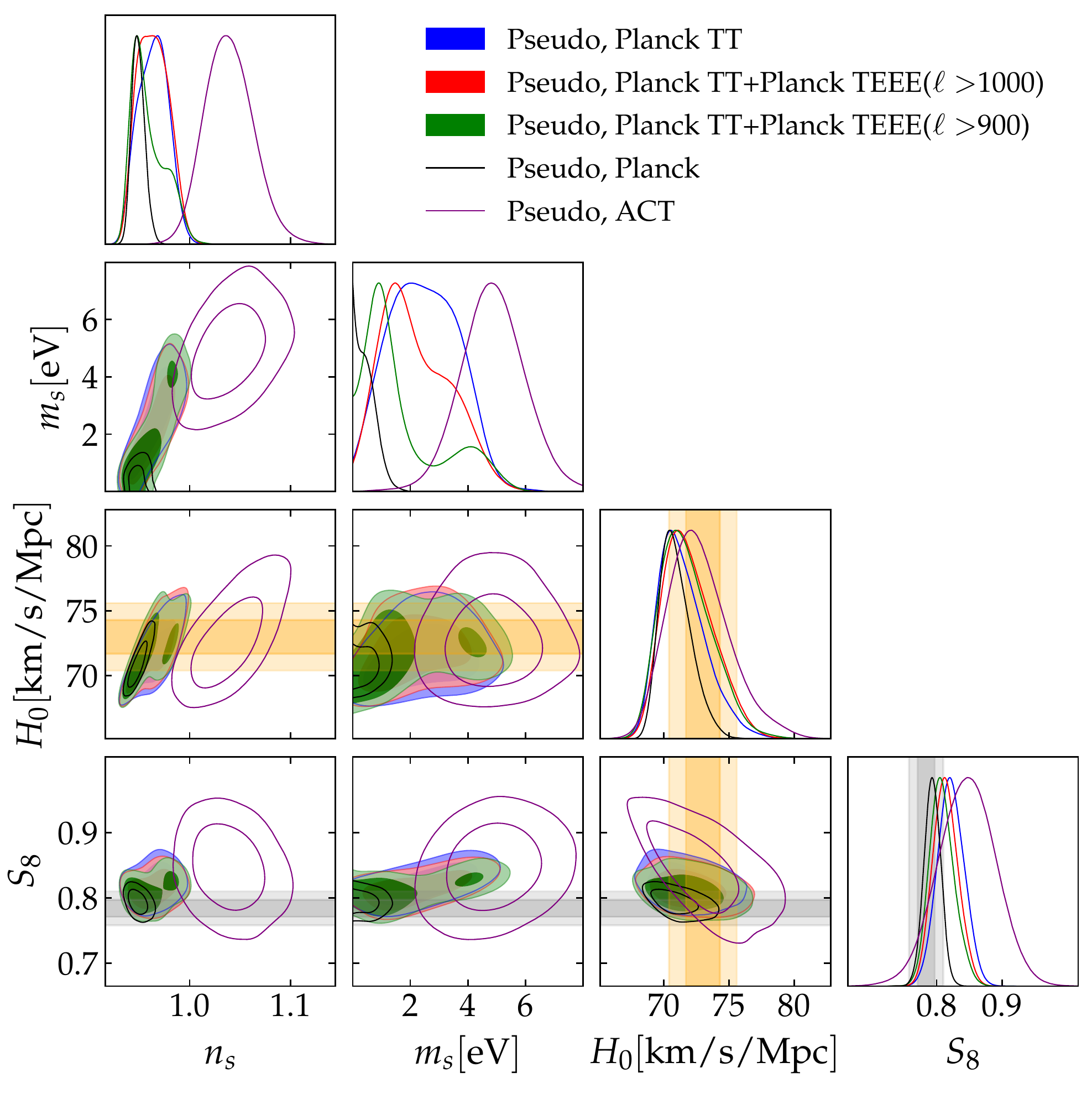}}
\caption{Posterior distributions of the cosmological parameters in the pseudoscalar (Pseudo) model for different subsets of \emph{Planck} TE-EE data. The orange and gray bands represent the direct measurements (1 $\sigma$ and 2 $\sigma$ confidence regions) of $H_0$ and $S_8$, from cosmic distance ladder~\cite{Riess:2020fzl} and weak lensing observations~\cite{Gatti:2021uwl}, respectively.}
\label{fig:Planckcut_Appendice}
\end{figure}

In Sec.~\ref{sec:act2}, we stated that the bound on the sterile neutrino mass comes from \emph{Planck} TE-EE data from $350\lesssim\ell\lesssim1000$. In this Appendix we discuss some of the additional analyses that we have performed to identify this multipole range. In Fig.~\ref{fig:Planckcut_Appendice} we examine the impact of different cuts in \emph{Planck} polarization data. While restricting \emph{Planck} TE-EE data to $\ell<350$ and $\ell>1000$, combined with \emph{Planck} TT, makes the analysis in statistical agreement with ACT, this is no longer true when \emph{Planck} TE-EE are restricted to $\ell<500$ or $\ell>900$.

\section{The CMB lensing anomaly in the non-interacting sterile neutrino model} \label{App:sterile_Vanilla}

 \begin{figure}[t]
 \centering
   \includegraphics[scale=0.4]{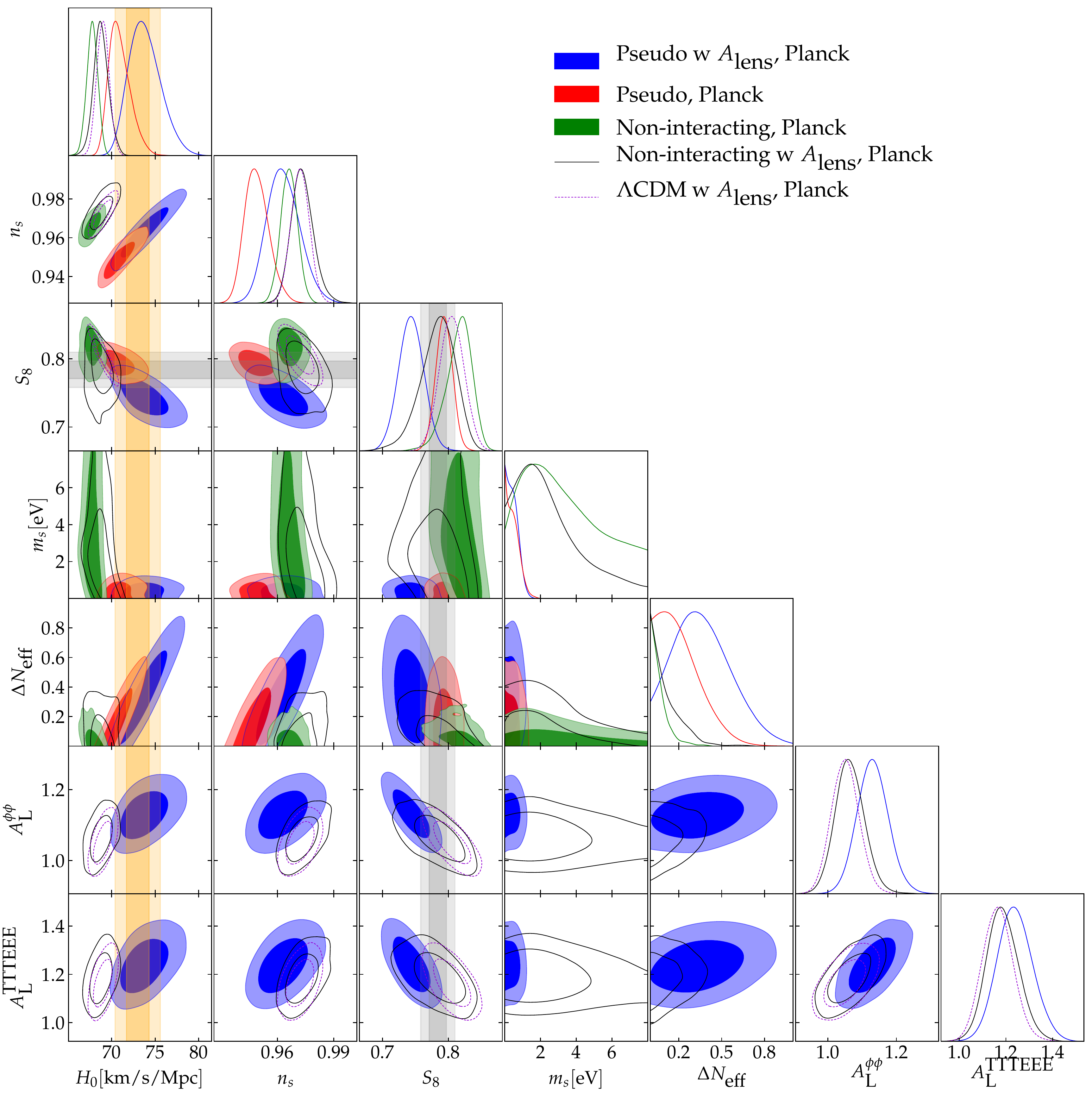}  
  \caption{\looseness=-1 Posterior distributions of the cosmological parameters in the pseudoscalar (Pseudo) and free-streaming model tested with \emph{Planck} data, with and without marginalizing over the gravitational lensing information. The orange and gray bands represent the direct measurements (1 $\sigma$ and 2 $\sigma$ confidence regions) of $H_0$ and $S_8$, from cosmic distance ladder~\cite{Riess:2020fzl} and weak lensing data~\cite{Gatti:2021uwl}, respectively. }
  \label{Fig: Pseudo_Vanilla_Alens_planck}
\end{figure}

 \begin{figure}[t]
 \centering
   \includegraphics[scale=0.45]{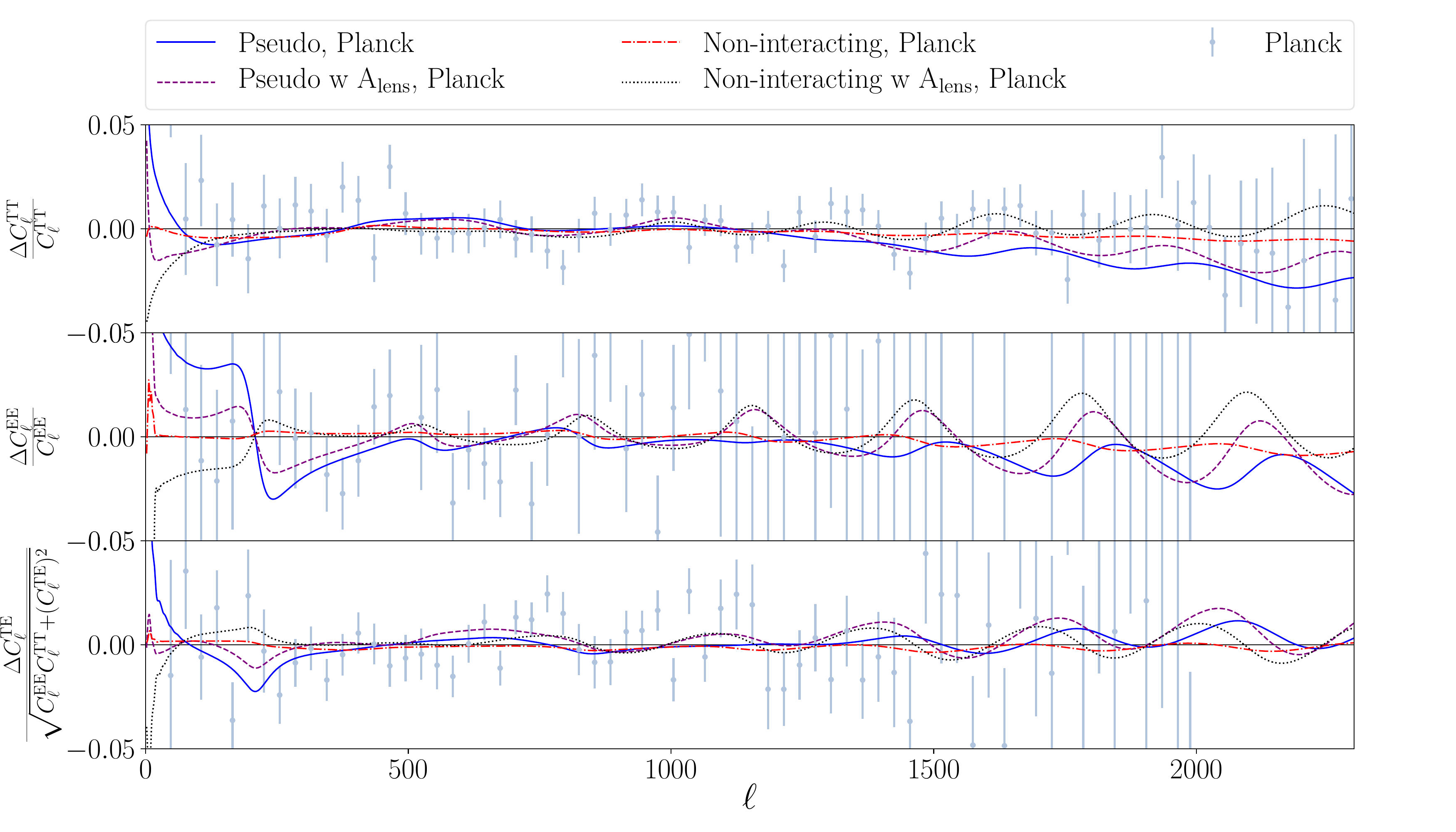}  
  \caption{Residuals of the pseudoscalar and non-interacting models in the CMB TT, EE and TE power spectrum with respect to the \emph{Planck}-only $\Lambda$CDM best-fit.
  We also show \emph{Planck} data-points and error bars. }
  \label{fig: ResidualVanilla}
\end{figure}

The goal of this Appendix is to quantify the impact of marginalizing over the lensing information in $Planck$ data in the presence of a non-interacting and free-streaming sterile neutrino species, and to compare the results with the ones discussed in Sec.~\ref{sec:Planck+Alens} for the pseudoscalar model. To this end, in Fig.~\ref{Fig: Pseudo_Vanilla_Alens_planck} we show the 1 and 2$\sigma$ confidence regions for the non-interacting case.
We note that the lensing anomaly is not relaxed even within such a scenario. We find that, as in the pseudoscalar case, in the non-interacting model the bounds on the sterile neutrino sector are not significantly modified by the introduction of the lensing parameters: 
a non-zero sterile neutrino mass is still allowed only in combination with $\Delta N_{\rm eff}$ values very close to zero.
Moreover, as already discussed throughout the paper, the introduction of the lensing parameters non-trivially modifies the degeneracy between cosmological parameters, such that higher values of $n_s$, $H_0$ and $\Delta N_{\rm{eff}}$ are predicted when $A^{\rm TTTEEE}_{\rm{L}}$ and $A^{\phi\phi}_{\rm{L}}$ are let free to vary -- albeit this trend is weaker than that observed in the pseudoscalar scenario (see Sec.~\ref{sec:Planck+Alens}). The impact on the CMB angular power spectra due to the preference for a higher value of $n_s$ is shown in Fig.~\ref{fig: ResidualVanilla}, where we compare the residuals of the pseudoscalar and the non-interacting best-fit power spectra with respect to the $\Lambda$CDM best-fit from \emph{Planck} data alone, with and without marginalizing over the $A^{\rm TTTEEE}_{\rm{L}}$ and $A^{\phi\phi}_{\rm{L}}$ parameters. In fact, as expected, we see that both in the non-interacting and in the pseudoscalar model, the best-fit spectra corresponding to the analysis with free lensing parameters feature more power on high multipoles, particularly in the TT-spectrum.

\section{Impact on the linear matter power spectrum}

 \begin{figure}[tb]
 \centering
 \includegraphics[scale=0.6]{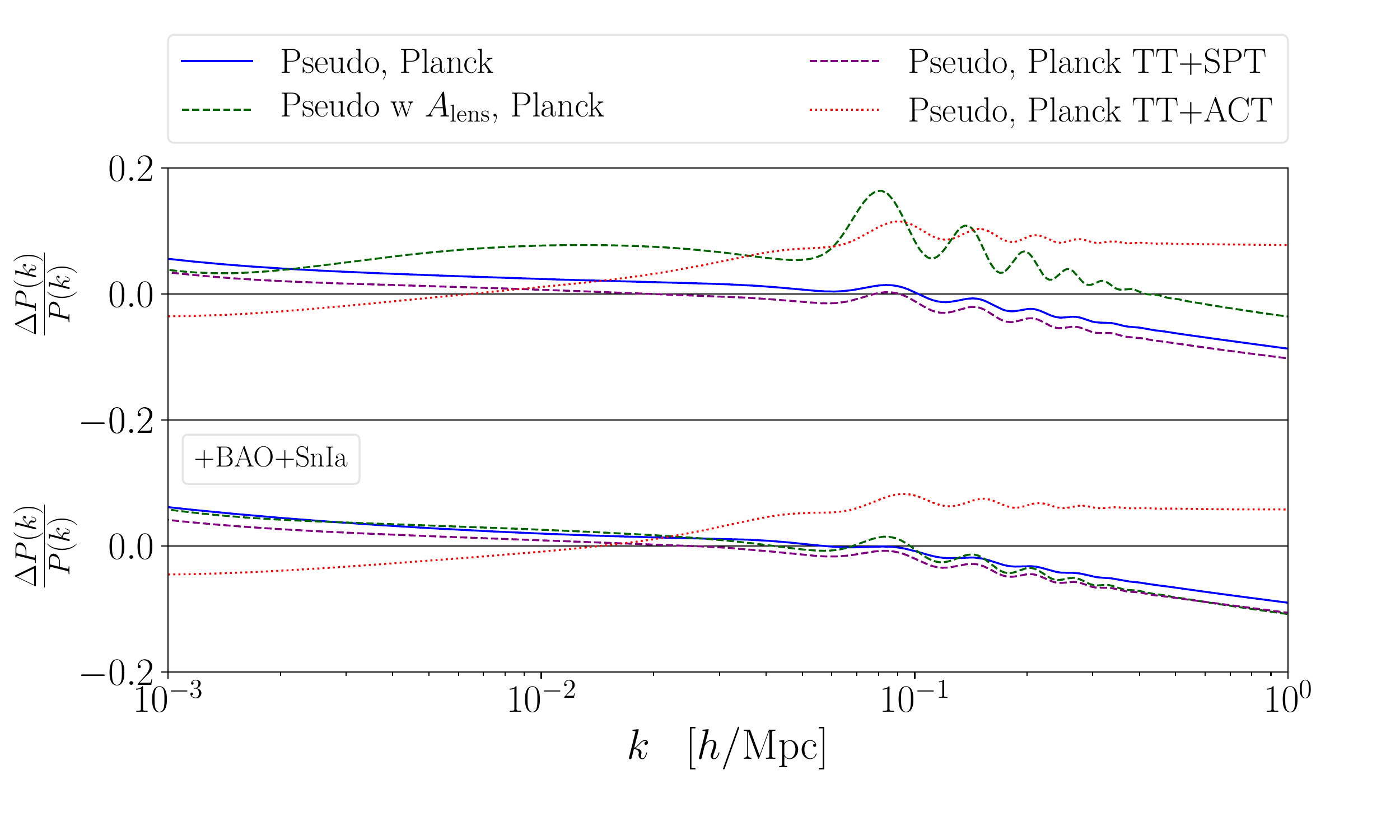}
 \caption{Linear matter power spectrum with residuals. For each model we show the relative difference between the pseudoscalar best-fit and the \emph{Planck}-only $\Lambda$CDM best-fit.}

\label{fig: ResidualMPS}
\end{figure}

In Fig.~\ref{fig: ResidualMPS} we show the best-fit residuals in the linear matter power spectrum with respect to the \textit{Planck}-only $\Lambda$CDM best-fit, for the pseudoscalar model tested against \emph{Planck}, \emph{Planck} TT+SPT, \emph{Planck} TT+ACT. In the upper  panel we show the CMB-only cases, while in the lower panel we report the cases where BAO and SnIa are added to the analyses.
As extensively discussed in Sec.~\ref{sec:model}, the pseudoscalar model does not induce a significant departure from the late-time matter distribution predicted by the $\Lambda$CDM model. The \emph{Planck} TT+ACT prediction is the only case featuring an enhancement rather than a suppression of power, driven by the higher value of $n_s$, and responsible for the higher $S_8$ value obtained in this analysis.
Finally, a comparison between the ``\emph{Planck}+$A_{\rm lens}$'' residuals in the two panels visually shows why the inclusion of BAO and SnIa data sensibly alters the constraints on that case only, as discussed in Sec.~\ref{sec:Planck+Alens}.

\bibliographystyle{JHEP}
\bibliography{references}

\end{document}